\documentclass[a4paper]{jpconf}
\usepackage{graphicx}
\usepackage[T1]{fontenc}
\usepackage[latin9]{inputenc}
\usepackage{amssymb}
\usepackage{esint}
\usepackage{hyperref}

\begin{document}
\title{Infrared behavior of Weyl Gravity}

\author{Les\l{}aw Rachwa\l{}  ${}^{1*}$ and Stefano Giaccari ${}^{2}$}

\address{1. Departamento de F\'{\i}sica -- ICE, Universidade Federal de Juiz de Fora, MG, Brazil}
\address{2.  Department of Sciences, Holon Institute of Technology (HIT), 52 Golomb St., Holon 5810201, Israel}

\ead{grzerach@gmail.com, stefanog@hit.ac.il}

\hspace{2.0cm}* - speaker at IARD 2020

\begin{abstract}
In this paper, we introduce and motivate the studies of Quantum Weyl Gravity (also known as Conformal  Gravity).
We discuss some appealing  features of this theory both on classical and quantum level. The construction of the quantum theory is described in detail to the one-loop level. To facilitate computations we use only physical degrees of freedom, which are singled out through the York decomposition. At the one-loop level we compute the partition function around a general Einstein space. Next, the functional renormalization group of couplings in Quantum Weyl Gravity is investigated. We reproduce completely previous results obtained on maximally symmetric and Ricci-flat backgrounds. Finally, we comment on further directions  and on the issue of conformal anomaly.
\end{abstract}

\section{Introduction}

Conformal gravity~\cite{Bach1} is a field-theoretical system with a very peculiar
form of relativistic dynamics. The evolution of Green functions in
this model is governed by two principles of the fundamental character.
First is the principle of general relativity since this model describes
gravitational interactions. Therefore we have the freedom to choose
an arbitrary coordinate system to describe the physics and nothing observable
depends on this choice. The second principle is the invariance under
local conformal transformations since this is a conformal gravity,
where the conformal symmetry is realized in the local version~\cite{Kaku}. In
turn, again we have a freedom to choose an arbitrary system of units
and scales to measure lengths, times, energies, etc. In conformal
gravity nothing observable depends on this choice of mass scale. When
these two principles are realized, the resulting dynamical system
is of a very special nature.

Relativistic dynamics is a very broad topic. In general, it describes
theories where the fundamental dynamical principles
governing the dynamics of the system are Lorentz invariant. When the Lorentz symmetry is
made local (gauged) then we end up with theories describing the dynamical
structure and evolution of spacetime itself. The requirement of local
Lorentz invariance puts very strong dynamical constraints on the possible
form of the relativistic theory of gravitation. Similarly, by gauging
conformal symmetry we produce a very particular gravitational theory,
which is unique in four-dimensional spacetime. The symmetry in the
global version describes matter systems and models which are in particular scale-invariant,
so they do not possess any characteristic invariant mass scale and
their dynamics looks the same at any scale of observation. In conformal
gravity we describe the dynamics of the relativistic gravitational
field which does not contain any mass scale (like Planck scale for
example). This situation is very special regarding gravitational theories
and one can realize that new additional constraints are put on the dynamics
of the gravitational field. At the end, the system is very constrained,
but not contradictory and instead it leads to unique very restrictive
predictions for gravitational observables. This is why it is interesting
to investigate this relativistic system closer.

Although conformal symmetry was first used by Weyl in what would be
called now a gravitational context, it was rediscovered in models
of particle physics in the second half of the last century~\cite{YM}. It was easier to find conformal high energy models without gravity than
with it. Conformal symmetry as well as scale invariance was first better understood
in massless and scaleless models of particle physics. In the realm
of quantum field theories (QFT) some very special models show conformality
also on the quantum level~\cite{Brink:1982wv}.  Since then this symmetry has been realized and investigated
both on the classical and quantum level of matter models~\cite{Col,Ad1,Ad2,Zee,Sp,K-S,Cap,Shaposhnikov}. However,
somehow in parallel, conformal methods found their applications in
classical research on gravitational theories~\cite{JKS,Ayazi,Ishiwata,Man1,Man2,Man3}. This put back conformal
symmetry to its natural gravitational setup, where, however,
it was most times analyzed in sufficient details only in the classical domain.
As it is well known, the question concerning the definition of a consistent quantum gravitational
theory is still open although there are very interesting proposals
on the market nowadays. The quantum role of conformal symmetry in
the gravitational setup is currently being actively investigated~\cite{Strominger,Hartnoll,Hill,Maggiore,JLK,Percacci1,Cornwall,Holdom}.
The constraints mentioned above select a unique theory in $d=4$ dimensions
which bears the name of Weyl gravity or conformal gravity (although
most of the preliminary works in this theory were done by Bach in
1920's~\cite{Bach1}).

Quantum Weyl Gravity (QWG)~\cite{Tseytlins} is a proposal for a fundamental theory realizing
symmetries of relativity and conformality in the local version both
on the classical and quantum level. We will discuss the status of
this theoretical model, its overall consistency and its current problems
below a bit. Most of them are related to the ultraviolet (UV) regime of
energies. One can assume hypothetically that they are solved by an
existence of some UV fixed point (FP) realizing also tentative UV-completion
of the theory. The situation on the other side of the spectrum, namely
in the infrared (IR) part of energies, is also interesting. In this
contribution we review the results obtained previously for the IR
renormalization group (RG) flows in QWG. We also put them in a perspective
relating them with a very special relativistic dynamics happening
in QWG.

\section{Motivations for QWG}

First, we discuss here the motivations for Weyl gravity (WG) on the
classical level. Later we add also some special features visible only
on the quantum level. Both particle models (without gravity) and gravitational
models with conformal symmetry are discussed briefly.

The first motivation for conformal symmetry in our Universe comes from
the rough observation of cosmic microwave background radiation (CMB).
Its spectrum (giving us some clues about the dynamics of the very
early universe) is very close to be scale-invariant~\cite{Ade}. The spectral
index of CMB is \begin{equation}
n_{s}=1.00-\varepsilon\quad{\rm with}\quad\varepsilon\ll1.\end{equation}
Hence as a first approximation we can assume that early Universe is
described by a scale-invariant gravitational theory coupled also to
scale-invariant matter models (like electromagnetic radiation). As
it is well known Maxwell theory of electromagnetic interactions is
not only globally scale-invariant, but is invariant under full local
conformal transformations (depending on the spacetime location). The
same extension can be performed on the level of gravitational dynamics
to end up with $4$-dimensional Weyl gravity. Moreover, it is also
a common wisdom that in a fully quantized theory involving relativistic
and dynamical gravity, all global symmetries must be made local. And
this procedure when applied to scale-invariance results in full invariance
under local conformal transformations. So the early Universe must
be described by a theory enjoying local (gauged) conformal dynamical
symmetry. This is one of the first phenomenological or observational
motivation for WG.

Conformal symmetry has its deep roots in the differential geometry
of two-dimensional curved surfaces. There it was first observed that
the following transformations\begin{equation}
g_{\mu\nu}\to g_{\mu\nu}'=\Omega^{2}(x)g_{\mu\nu}\label{eq: conftrans}\end{equation}
make any curved $2$-dimensional manifold flat. (This is why all $2$-manifolds
are conformally flat). The transformations in (\ref{eq: conftrans})
are not induced by any coordinate transformation $x\to x'=x'(x)$,
hence the differential structure of the manifolds is not preserved.
Under the conformal transformations the metric is ``charged''
whereas the coordinates remain untouched, so that  the invariant
infinitesimal line element in differential geometry is not preserved
and scales according to the law\begin{equation}
ds^{2}\to ds'^{2}=\Omega^{2}(x)ds^{2}.\end{equation}
From this we derive that distances (lengths, times, energies, etc.)
are not invariant. They are relative. And they depend on the choice
of the gauge; in this case we speak about a conformal gauge. We conclude
that by performing conformal transformations we abandon the absolute meaning
of physical distances and scales \cite{Rachwal:2018gwu}. With physical relativity we know
that velocities are relative, here we add the conformal gauge-dependence
of scales. But one element of the geometrical description remains invariant
under the conformal transformations, even when their parameter $\Omega=\Omega(x)$
depends on the location on the manifold. These are angles between
vectors. The angle between two vectors $\vec{u}$ and $\vec{v}$ could
be defined as\begin{equation}
\measuredangle(\vec{u},\vec{v})=\arccos\left(\frac{\vec{u}\cdot\vec{v}}{\sqrt{|\vec{u}|^{2}|\vec{v}|^{2}}}\right).\end{equation}
This is an example of conformal invariant since it does
not depend on the scales.

In Minkowskian framework (when we include temporal relations to our
differential geometry description of curved spacetime), the structure
which is left invariant under conformal transformations is a causal
structure on the spacetime manifold. The light-like trajectories of
massless particles remain such since their defining condition $ds^{2}=0$
is untouched by conformal rescalings (\ref{eq: conftrans}). This
was one of the reasons for the success of methods of conformal transformation
during the golden years of general relativity (GR). In the gravitational
framework, conformal methods revealed secrets of black holes and their
horizons as well as they were instrumental in flattening of cosmologies.
(All Friedmann-Lema\^i{}tre-Robertson-Walker gravitational spacetimes
are conformally flat.) Only the causal or conformal structure of the
spacetime manifold is preserved after the process of conformal change
of the metric. Moreover, in the gravitational context global scale
symmetry (equivalent to absence of mass dimension parameter in the
theory, so all the couplings are dimensionless) can be gauged like
 other global gauge symmetries, which can be made local and dynamical, for
example in some models of particle physics. For the gravitational
dynamics, the local conformal symmetry (invariance with respect to
transformations (\ref{eq: conftrans}) with the spacetime-dependent parameter
$\Omega(x)$) can be an additional gravitational symmetry constraining
the dynamics even more strongly, as we will show below (both on the quantum
and classical level). The conformal symmetry can be treated like a
gauge symmetry of gravity and can be placed on the same level as local
Lorentz symmetry (so invariance under local boosts together with local translations)
in the full gravitational theory.

Conformal symmetry should be the last dynamical symmetry (making
impacts on the dynamics of the system) to be discovered in particle
physics. For its verification one needs to go to arbitrarily high energy
scales. However, the first time this symmetry
was realized on the quantum level was in theoretical models without
gravity and on flat Minkowski spacetime backgrounds \cite{Brink:1982wv}.

In high energy models, one intuitively knows that almost every time
one deals with massless particles or excitations, then the conformal
symmetry emerges in the description of the system. Even if the system
is not intended to possess some conformal properties, the limit where
all particles become effectively without mass, leads to a miraculous
enhancement of the symmetries of the considered model. And there are some
properties that could be only explained by invoking conformal symmetry.
Hence conformal symmetry is a quite frequently an emergent approximate
symmetry of models of QFT, when the regime of high energy is analyzed.
The particles do not have to be strictly with vanishing mass, their
mass can be effectively very small compared to other energy scales
in the very high energy approximation. For the classical considerations
done on flat Minkowski spacetime background, conformal symmetry can
be seen just as global scale-invariance. However, full conformal group
contains more generators than just dilatation generator from scale-invariance.
And also its predictions are more firm and it constrains the dynamical
system enjoying invariance under the full conformal group more.

The full constraining power of conformal symmetry is seen on the Green
functions of the model enjoying it. It is more than just scale-invariance,
which forbids only the presence of dimensionful parameters in Green
functions. For global scale-invariance dimensionful momenta $p$,
energies $E$, and fields $\phi$ can still be present on the legs
of Green functions and can enter into the corresponding expression
for these Green functions in arbitrary combinations not constrained
by any further symmetry principle. Full conformal symmetry changes
this last fact. The dynamics of a Quantum Field Theory (QFT) with conformal symmetry at
quantum level is so strongly constrained that  it is customary to refer to this special class of theories  under the name of conformal field theories (CFT). Quantum Green
functions are very strongly constrained in such models. For example, the form
of all $2$-point and $3$-point functions is completely fixed and
$4$-point functions at arbitrary loop order depends only on
one arbitrary function of very specially constructed conformal ratios~\cite{Rychkov:2016iqz}. Quantum
conformality enjoyed by the theories in this class means that all
quantum fluctuations look the same at all energy scales. So we do
not have standard suppression of quantum fluctuations in IR or in
UV. In quantum CFT models they show the same strength in every energy regime.
But quantum conformality is more than quantum scale-invariance since
these two group of symmetries are not isomorphic to each other. Moreover,
as a next piece of motivation, one can notice that the current standard
model (SM) of particle physics is a classically almost scale-invariant
theory and the breaking of scale-invariance is related to the spontaneous
breaking of electroweak symmetry (so it is proportional to the vacuum
expectation value of the Higgs field, or in turn to its mass).

Relativistic models with scale-invariance enjoy a symmetry
under an $11$-parameter group, which includes the generator of dilatations, apart from those from the Poincar\'e{} group. Instead, full conformal group in $d=4$
contains $15$ generators because of $4$ additional generators of
conformal boosts. To require invariance under these additional transformations
of all Green functions (also on the quantum level) is a big constraint,
which signifies that there is an additional symmetry on the quantum
level. This is of course also a symmetry of the full quantum effective
action. Conformal symmetry as additional symmetry on the quantum level
constrains further (besides relativistic constraints imposed by symmetry
under transformations from the Lorentz group) quantum dynamics in
such a way that UV-divergences are completely avoided in such models.
Since there cannot be any scale or mass generated in quantum dynamics
of CFT model, then there is no need to renormalize the quantum theory
by introducing an arbitrary renormalization scale $\mu$. And the
fact of need for doing infinite UV-renormalization is tightly related
to the presence of infinities in the UV regime of the theory. This
additional symmetry is essential for avoiding quantum singularities
in CFT models. However, the similar arguments can be also applied
to classical gravitational theory enjoying conformal symmetry to solve
the ubiquitous problems with classical spacetime singularities. This
is a next motivation for classical Weyl gravity, which should keep
its conformal properties also on the quantum level for this argument
to work.

The CFT matter models considered above are indeed very special QFT's
since in them there are no UV-divergences. This implies that there
are no $\beta$-functions both on the perturbative (to all orders)
level as well as non-perturbatively due to conformal algebra in which
we must have the energy-momentum tensor operator of CFT. If we do
not need to use the renormalization scale $\mu$ to define properly
theory on the quantum level, then also the form of the quantum effective
action is very special. One can view the quantization procedure ${\cal Q}$
as a several mapping in theory space ${\cal T}$:\begin{equation}
{\cal Q}:{\cal T}_{{\rm cl}}\to{\cal T}_{q},\end{equation}
where both the classical theory ${\cal T}_{{\rm cl}}$ and the quantum
theory ${\cal T}_{q}$ (regularized and renormalized with some physical
renormalization conditions) are elements of the theory space ${\cal T}$.
The quantum theory ${\cal T}_{q}$ is specified completely by quantum
effective action $\Gamma$ (regularized and renormalized), similarly
like ${\cal T}_{{\rm cl}}$ is defined by classical action functional
$S_{{\rm cl}}$. The conformal symmetry constrains the effective action
of the quantum theory ${\cal T}_{q}$ so much that ${\cal T}_{q}={\cal T}_{{\rm cl}}$
(up to a finite renormalization of couplings). In other words, the
quantization procedure ${\cal Q}$ finds its mathematical fixed point
for very special initial elements ${\cal T}_{{\rm cl}}$, which leads
to conformal symmetry both on the classical level. And what is more
important it leads also to conformality on the quantum level which
is preserved by quantum corrections. In a mathematical sense this
gives a fully idempotent definition of the quantum effective action
$\Gamma$ in CFT. Quantization process ${\cal Q}$ produces quantum
corrections to the classical dynamics, but they are scale-invariant,
so their dynamics is completely captured by the original classical
action, if it is properly conformally invariant. We conclude that
quantum effective action $\Gamma$ is the same (up to finite dimensionless
coupling rescalings) as the classical conformally invariant action
$S_{{\rm cl}}$.

The conformal symmetry of CFT is explicitly realized on the form of
all correlation functions of the theory. But not only there. Due to
conformal symmetry the special set of Ward identities is satisfied
in CFT relating different $n$-point function of the theory. They
are quantum operatorial manifestation of the conformal symmetry present
in the quantum effective action $\Gamma$. They encode on the quantum
level, for example, the fact that the trace of the effective energy-momentum
tensor $T=g^{\mu\nu}T_{\mu\nu}$ of the full theory is off-shell zero
even after inclusion of quantum corrections. And this statement has
the operatorial meaning, not only in the sense of expectation values.
Generally, the set of all conformal Ward identities satisfied by any
CFT (including these models when gravity is dynamical) is an expression
of the fact that the quantum system is highly constrained and that
its relativistic dynamics is not an ordinary one. Moreover, quantum
CFT models are finite (no UV divergences), so these are very well
behaved quantum models which do not need regularization for their
definitions on the quantum level. The conformal symmetry in CFT's
is very constraining such that the form of classical action describing
the starting point of the theory $S_{{\rm cl}}$ is very restrictive,
provided that the number of spacetime dimension $d$ is fixed. Another
specification of the model is the listing of all fields in it. However,
not any composition is allowed. There are also restrictions on this.
For example, if we desire too construct a CFT with gauge interactions,
this must take a form of a very special theory of ${\cal N}=4$ supersymmetric
Yang-Mills theory. If we want to add also gravitational interactions,
this must be a model of ${\cal N}=4$ conformal supergravity first
discovered in 1985 by Fradkin and Tseytlin. Other actions, when conformal
symmetry is preserved on the quantum level (so it is not broken) are
not possible in $d=4$.

We see that in CFT there are no $\beta$-functions of (necessarily)
dimensionless coupling parameters. So they are really coupling constants,
in the old sense, they are not RG running constants. There is no RG flow in models of CFT. The only difference from the classical
field theory is that the dimension of operators may receive some finite
corrections (though not related to any RG running), so there could
be some anomalous dimensions of various operators in CFT's. The fact
that there is no RG flow is equivalent to the statement that conformal
field theories sit at the FP of the RG flows. They describe completely
the situation (scaling dimensions, also known as critical exponents)
of all operators present in a given CFT. For a different point of view, see \cite{Codello:2012sn}.
The analysis of FP's of RG
flows is a very important topic for its own interest. However, knowing
the general structure of CFT's in a form of CFT data (conformal algebra,
set of primary operators, their scaling dimensions) helps a lot in
describing the FP's of RG flows which are very special points in theory
space ${\cal T}$. As it is well known FP's of RG are starting (or
ending) points of RG flows since the situation at the FP's could be
seen as very extraordinary. But the existence of non-trivial RG flows
breaks explicitly conformal symmetry of the CFT model. To do this
in a more gentle manner, one can imagine turning smoothly various
deformation operators and then the breaking of conformal symmetry
could be achieved in a spontaneous way. In this way CFT's are starting
points in the theory space ${\cal T}$ for various non-conformal theory
modifications, in which there is RG flow, but still there is some
remnant and memory of the constraints on the dynamics that were imposed
at the FP. Therefore this kind of relativistic dynamics enjoys RG
flows and only some remnants of full conformal symmetry. The full
symmetry is recovered in the UV limit of the theory. In the next sections
of this contribution, we will present one quantum computation which
is done in the theory, which originates from full CFT with gravitational
interactions, but in which conformal symmetry is spontaneously broken
and the RG flows towards IR is possible. As we will see even in such
partially broken model, the amount of constraints which were left
after fully realized conformal symmetry is significant and the ensuing
relativistic dynamics is constrained and highly interesting.

\section{Weyl gravity}

Now we proceed with the construction of Weyl gravity in $d=4$ spacetime
dimensions. The number of dimensions here has to be fixed and even
integer. In quantum theory we will not allow playing with the ``effective''
dimensionality of spacetime similarly to what happens for example in dimensional
regularization (DIMREG) scheme). This requirement has to do with the
construction of the classical action of gravity which is supposed
to be conformally invariant. For such a construction we must use Weyl
tensors as building blocks, and if we require locality of the action,
then they can appear only in positive integer powers. Hence the dimension
$d$ of spacetime has to be even integer, and we choose $d=4$ to
be consistent with our every day experience of relativistic physics.

The main building block for the action is the Weyl tensor (tensor
of conformal curvature), which is defined in $d=4$ as\begin{equation}
C_{\mu\nu\rho\sigma}=R_{\mu\nu\rho\sigma}-2R_{[\mu[\rho}g_{\sigma]\nu]}+\frac{1}{3}g_{[\mu\rho}g_{\nu]\sigma}R,\label{eq: weyldef}\end{equation}
where $R_{\mu\nu\rho\sigma}$, $R_{\mu\nu}$ and $R$ denote the Riemann
tensor, Ricci tensor and Ricci scalar respectively. Its main useful
property is the conformal transformation law\begin{equation}
C_{\mu\nu\rho\sigma}\to\Omega^{2}C_{\mu\nu\rho\sigma}\end{equation}
or for versions with different positions of indices (such position
matters!)\begin{equation}
C_{\mu\nu\rho}{}^{\sigma}\to C_{\mu\nu\rho}{}^{\sigma},\quad{\rm and}\quad C^{\mu\nu\rho\sigma}\to\Omega^{-6}C^{\mu\nu\rho\sigma}\end{equation}
Of course, Weyl tensor plays the role of the gravitational curvature and
for example the vanishing of all its components for some given spacetime  implies it is conformally flat. Then such a spacetime can be mapped
to a completely flat one by a conformal transformation (\ref{eq: conftrans})
of the metric tensor. Generally, in conformal gravity different spacetimes
differ only by their conformal structures (so how the Weyl tensor
$C_{\mu\nu\rho\sigma}$ looks like), and the details about the differential
metric structure are irrelevant. Moreover, metrics related by conformal
transformations are in the same conformal equivalence class and hence
they describe the same gravitational spacetime from the point of view
of conformal symmetry. Not only we have to be independent of the particular
choice of the coordinate system, but also of the choice of the conformal
gauge, so of the rescaling factor $\Omega^{2}$ in the definition
of conformal transformation for metric (\ref{eq: conftrans}). This
conformal factor $\Omega^{2}$ (known also as a scale factor in cosmology)
is strictly non-negative due to the usage of the square of a completely
arbitrary real scalar function $\Omega(x)$ of the point $x$ of the
spacetime. To describe the gravitational field in conformal gravity,
one must exploit conformal invariants, which are both diffeomorphism
invariants (scalars) as well as they do not transform under general
conformal transformations. The set of them is quite restricted. The
gravitational action must be one of them.

This action~\cite{Bach1} in $d=4$ is constructed exclusively with the help of
Weyl tensor as\begin{equation}
S_{{\rm conf}}=\!\int\! d^{4}x\sqrt{|g|}\alpha_{C}C^{2},\label{eq: weylgravaction}\end{equation}
where the square of the Weyl tensor has the following expansion in
known quadratic in curvatures invariants\begin{equation}
C^{2}=C_{\mu\nu\rho\sigma}C^{\mu\nu\rho\sigma}=R_{\mu\nu\rho\sigma}R^{\mu\nu\rho\sigma}-2R_{\mu\nu}R^{\mu\nu}+\frac{1}{3}R^{2},\label{eq: c2expansion}\end{equation}
which shows a big resemblance to the formula (\ref{eq: weyldef})
regarding the numerical coefficients. This is easily explained by
the tracelessness of the Weyl tensor in any pair of its indices. To
check that the action (\ref{eq: weylgravaction}) is conformally invariant
in $d=4$ spacetime dimension, one must also recall a conformal transformation
law for the volume element, namely\begin{equation}
\sqrt{|g|}\to\sqrt{|g'|}=\Omega^{4}\sqrt{|g|},\end{equation}
while the coordinate volume differentials collected in $d^{4}x$ do
not transform (that is $d^{4}x\to d^{4}x$) since the coordinates
are not changed under the transformation ($x^{\mu}\to x^{\mu}$) and
all conformal transformations are induced from the way how the metric
changes in (\ref{eq: conftrans}).

One notices that the scalar invariant $C^{2}$ transforms co-covariantly,
that is we have\begin{equation}
C^{2}\to\Omega^{-4}C^{2}\end{equation}
and only its properly densitized version\begin{equation}
\sqrt{|g|}C^{2}\to\sqrt{|g|}C^{2}\label{eq: densitizedinv}\end{equation}
is conformal invariant. However, this last one is not a scalar invariant
from the perspective of GR, since it is a density. Therefore in the
local framework we are in conflict of finding local conformal invariants
which are also scalars from the point of view of diffeomorphisms.
This observation has some important consequences for the study of
black hole solutions and the question of the fate of singularities
there. Probably, one should look at some non-local invariants which
are integrated versions of the densitized scalar invariant in (\ref{eq: densitizedinv}),
i.e. at the values of the action integral (\ref{eq: weylgravaction})
over some small region of spacetimes.

The action integral is a scalar and it is conformally invariant, but
as a main feature of the action functionals it is obtained via integration,
so that is why it must be highly non-local (or strongly non-local).
This action in (\ref{eq: weylgravaction}) describes completely the
classical conformal Weyl gravity in $d=4$. This theory is both diffeomorphically
(relativistic coordinate changes) and conformally invariant (changes
of the conformal factors of the metric). This is a unique theory in
$d=4$ since there are no other local conformal invariants (besides
$\sqrt{|g|}C^{2}$ in (\ref{eq: densitizedinv})) possible to be constructed
here. Additionally, the theory is described by only one conformal
coupling constant $\alpha_{C}$. In this way this theory is a very
similar to Maxwell or Yang-Mills action, which is also quadratic in
gauge field strengths and which also comes with one dimensionless
coupling constant. The fact that $\alpha_{C}$ is dimensionless and
unique constrains the choice of the theory even more in the gravitational
setting. Compare this situation with Einstein-Hilbert with cosmological
constant which already comes with two coupling parameters. Models
with higher derivatives and quadratic in curvatures contain even
more such coupling parameters. (The general Stelle theory has four
such parameters appearing in the Lagrangian $\lambda_{{\rm cc}}+\kappa_{N}R+\alpha_{R}R^{2}+\alpha_{C}C^{2}$).
The uniqueness of the theory is a very elegant and economic property
to select a correct gravitational theory. For definiteness in Euclidean
setup we require that the coupling $\alpha_{C}$ is given by a real
number and strictly positive.

One should examine the advantages of classical Weyl gravity as a theory
of relativistic gravitational field. First, it is noticed that the
energy-momentum tensor which is a source of gravity must be here traceless
to comply with the tracelessness of the tensor of equations of motion
(EOM) coming from the action (\ref{eq: weylgravaction}) required
by conformal invariance of both matter (source) and gravity sides
of the coupled EOM. Explicitly, the tensor of EOM for Weyl conformal
gravity has the form\begin{equation}
B^{\mu\nu}=\frac{1}{\alpha_{C}}\frac{1}{\sqrt{|g|}}\frac{\delta S_{{\rm conf}}}{\delta g_{\mu\nu}}=2\nabla_{\kappa}\nabla_{\lambda}C^{\mu\kappa\nu\lambda}+C^{\mu\kappa\nu\lambda}R_{\kappa\lambda}\label{eq: bacheom}\end{equation}
and it bears the name of Bach tensor. Generally, in $d=4$ the Bach
tensor is symmetric ($B^{\mu\nu}=B^{\nu\mu}$), automatically traceless
($g_{\mu\nu}B^{\mu\nu}$), divergence-free ($\nabla_{\mu}B^{\mu\nu}=0$)
and it originates from the conformal action $S_{{\rm conf}}$ as the
variational derivative of the action integral in (\ref{eq: weylgravaction}).
The field equations in Weyl gravity read\begin{equation}
B^{\mu\nu}=\frac{1}{2}\alpha_{c}T^{\mu\nu}\end{equation}
This form of EOM puts a stringent condition on the form of a possible
energy-momentum tensor of the matter source because it has to be traceless,
so that only massless or scale-invariant matter on their equations of motion can give rise to it. This condition
is satisfied in the deep UV (regime of high energies), where all the
matter becomes {}``effectively'' massless and conformal symmetry
is restored. Since the effects of relativistic gravity are not well
investigated and understood in the regime of very high energies and
also of high matter densities, then it is more practical to look for
tests of CWG in the setup of vacuum solutions, where there is no matter
at all ($T^{\mu\nu}=0$).

Here, for empty space solutions, we can perform and verify three basic
tests of gravitational theory which originally gave impetus to
the Einstein gravitational theories. Those were the perihelion of
Mercury, the light deflection by the Sun and the redshift of spectral lines
due to gravitational effects. To this one could also add the time
delay of radar echoes and precessions of gyroscopes. All these tests
were carried out in empty space and they gave a verification of Einstein's
theory. But they are also equally valid and satisfied in CWG since
they are based in particular on static and spherically symmetric exact
vacuum solutions in the form of the Schwarzschild metric. This metric
is also an exact vacuum solution in CWG. A bolder statement is that
all exact vacuum solutions of Einstein gravitational theory (without
cosmological constant) are also exact vacuum solutions of classical
Weyl gravity. For this, one have to notice that in the expansion in
(\ref{eq: c2expansion}) only the Ricci tensor appears, hence the
EOM of CWG are quadratic in Ricci tensors or in Ricci scalars (and
not of Riemann tensor), and hence they vanish identically on Ricci-flat
spacetime which are known to be gravitational vacuum in original Einsteinian
theory. We conclude that metrics describing vacuum spacetimes like
Schwarzschild, Kerr, etc are also solutions there. This implies that
CWG reproduces all precise tests of relativistic gravitation like
Einstein theory in the vacuum.

But CWG contains more solutions that just Einsteinian theory. And
this is one of its powers. For example, the solutions found by Mannheim
and Kazanas for static and spherically symmetric situations in conformal
gravitational vacuum~\cite{Mannheim:1990ya} are a family of solutions with more parameters
than in Einsteinian theory. But they can serve as a solution of ``dark
matter'' or missing matter problems in galaxies~\cite{Man1,Man2,Man3}. These solutions
describe accurately more than 200 galactic rotation curves with just
one fittable parameter and mimic the behavior of the gravitational
potential  produced by the conjectured dark matter halo. However, in CWG dark
matter is not necessary at all and the power of classical conformal
gravity is to explain flat rotation curves as exact vacuum solutions
of the theory. Moreover, as emphasized at the beginning of the previous
section conformal Weyl gravity provides also an explanation for ``almost''
scale-invariant power spectrum of scalar cosmological fluctuations
therefore explaining one of the issues in current cosmology. Other
problems in cosmology where conformal gravity helps is the problem
of cosmological constant, which is not needed in conformal cosmology
any more, and the problem of flatness. According to ~\cite{Mannheim:2005bfa,Mannheim:2020lfz}
conformal gravity also solves the problem related to inflation and
dark energy problems (vacuum energy) since all cosmological manifolds
are conformally flat. This fact also simplifies the analysis of the
cosmological fluctuations both in the scalar as well as in the tensorial
sectors.

Classical CWG provides also a natural and very robust resolution of
the problem of spacetime singularities, which are probably the most
serious inconsistencies of classical relativistic gravitational physics.
These singularities signal the breakdown of classical relativistic
physics where the theory inherently sets its own limits of validity.
One would say that the fate of classical singularities of the gravitational
field is to be eventually solved somehow by a hypothetical theory
of quantum gravity. However, we think that for the full resolution
of them, one has to use the power of additional new local symmetry
in the gravitational framework. And this symmetry is a dynamical symmetry
related to arbitrary conformal rescalings of the metric of the spacetime
(\ref{eq: conftrans}). By allowing for conformal rescalings one resolves
the problem of classical GR singularities (which put in danger exact
solutions of the gravitational theory). For example, the Big Bang
or Big Crunch singularities in the cosmological framework, or black
hole singularities (Schwarzschild, Kerr, and other metrics, which
are singular at the origin of the black hole)~\cite{narlikar:1977nf,Bambi:2016wdn,Modesto:2016max}. The idea is to relegate
the ``observable'' general relativistic singularity
into a (conformal) gauge-dependent unobservable conformal factor $\Omega^{2}$.
In this way something which was for example the curvature singularity
(like measured by the Kretschmann invariant scalar $R_{\mu\nu\rho\sigma}^{2}$)
in Einsteinian theory, now is not anymore a good conformal invariant
to look at. By performing a change of the conformal gauge according
to (\ref{eq: conftrans}), one shows that such an invariant scalar
from GR loses its unambiguous meaning and its actual value changes\@.

When one analyzes the conformal invariants (there are very few of
them), then one sees that the singularities can be completely resolved
and that the conformal equivalence class of metrics is not singular,
similarly like the coordinate ``singularity''
at the horizon of Schwarzschild metric is not present in the equivalence
class of metrics related by coordinate transformations. The point
is that there exist metrics (describing physical gravitational spacetimes),
which are not singular in these classes, compared to one particular
example metric which was singular but the singularity was assessed
by looking at ``wrong'' invariants, that
is scalars which are not invariant under the full group of symmetries
of the theory in question. Then the choice of non-singular metrics
should be a preferred one, and the one of original singular metric
should be discarded as driven by a singular conformal gauge-dependent
scale factor $\Omega^{2}$ of the metric. Since the choice of any
metric in the equivalence class is possible (and should be done to
perform actual physical computations), then the one not giving rise
to singularity is completely equivalent to others, but about the resolution
of singularities one decides by choosing a most convenient one from
the physical purposes. And that is why one choose Kruskal-Szekeres
coordinates for example over original Schwarzschild coordinates to
cover the whole black hole horizon region. One concludes that there
is not a true singularity at the location of the horizon, it was only
a coordinate gauge choice singularity. In complete analogy for the
conformal resolution of singularities, one derives that the conformally
related metrics which do not develop singularities are better and
physically more convenient than the original metrics in Einsteinian
theory. One concludes that there is not a true singularity with a
conformally invariant meaning here, it was only an original conformal
gauge choice that was singular and led to not ``conformally''
invariant notion of curvature singularities like in Einsteinian GR.
This again shows the interplay and similarities between relativistic
and conformal ways of treatment of some things - a theme which is
constantly recurring in this original contribution.

We also believe that the true resolution of classical singularities
still should be achieved on the classical level of the theory. Even
if the quantum effects play an important role for this issue, they all
are captured by the quantum effective action. The true quantum-corrected
solutions of the theory are understood as classical solutions to the
quantum effective action, so the same problem returns to the classical
level. Our take on is to assume that additional new symmetry in the
gravitational relativistic dynamics comes with the rescue to these
annoying problems of ubiquitous and inevitable classical singularities
of gravitational spacetimes.

Another virtue of CWG is that this approach to relativistic dynamics
of the gravitational field is fully Machian. This realizes completely
the old ideas of Mach, which originally inspired Einstein in his construction
of the gravitational theory. However, as it is well known in Einsteinian
gravity the dynamics is only partially Machian, as this is obvious
from Einstein equations in vacuum, which depend on the matter distribution
only locally. However, in CWG gravitational field at a particular
location is determined completely by the conformal matter distribution
everywhere else in the Universe (in Machian sense, although of course
the EOM of CWG are local despite that they are with four derivatives
on the metric tensor). This particular feature is seen also from the
explicit form of Mannheim-Kazanas solutions for which the particular
slope of the galactic rotation curve is determined from cosmological
reasons, so the matter density in the whole universe influences the
gravitational potential inside galaxies. In particular, Mannheim-Kazanas
solutions provide one exact solution to the problem of embedding the
local galactic gravity solution in a bigger scale picture of cosmological
expansion of the Universe as a whole. The fully Machian character
of CWG is obviously related to the preservation of only causal structure
of the gravitational spacetimes under conformal transformations. And
only the last structure of the spacetime manifold retains its meaning
in full CWG.

Conformal symmetry is quite often seen on the classical level of the
relativistic dynamics. For example, classical electromagnetic and
Yang-Mills (YM) (non-Abelian) gauge theories, and the theory of massless
Dirac fermions in $d=4$ spacetime dimensions, are perfectly conformally
invariant. They enjoy spacetimes symmetries from the full $15$-parameter
conformal group containing as its subgroup the $10$-element usual
Poincar\'e{} group of standard relativistic symmetries. Moreover,
a scalar field conformally coupled to gravity in $d=4$ with the non-minimal
coupling of the form\begin{equation}
S_{{\rm int}}=\!\int\! d^{4}x\sqrt{|g|}\frac{1}{6}R\phi^{2}\end{equation}
represents another classically conformal model with matter. One can
also ask what are the corresponding consequences of this fact on the
quantum level. For this one has to discuss the quantum Weyl gravity
QWG.

\section{Quantum Weyl Gravity}

There are various reasons for quantum Weyl gravity. More clear conceptually
is to couple first quantum matter fields to fixed but non-trivial
background geometry. This setup of quantum fields in curved spacetime
is well known and it has very profound consequences also for the construction
of resulting quantum gravitational theory, where the gravity is taken
as dynamical and the dynamics of the related field of the graviton
is fully quantum. One can induce gravity from quantum matter fluctuations~\cite{Visser:2002ew}.
In this way one can find a correct quantum gravity theory, the same
like this was originally done with YM theory when at the beginning
only fermions transforming non-trivially under non-Abelian symmetry
groups were known.

The idea is the following. It is the embodiment of the original deWitt-Utiyama
argument~\cite{Utiyama:1962sn}. When the matter fields are put in a non-trivial classical
external non-dynamical gravitational field, then as quantum fields
described by QFT on non-trivial backgrounds they produce various UV
divergences, both in the matter and in the gravitational (geometric
sector). The correct quantum gravitational theory should absorb all
the divergences of the gravitational character in its counterterms.
For this the original action of the gravitational theory should contain
the corresponding terms. And here the results of the computation about
these UV divergences is as follows. When one uses  the conformal models described above as matter models
on curved spacetimes, then in
$d=4$ spacetime dimensions the only non-trivial gravitational counterterm
which is generated by quantum corrections induced from the matter
side is of the conformal character, that is, it is precisely\begin{equation}
S_{{\rm div}}=\!\int\! d^{4}x\sqrt{|g|}C^{2}.\end{equation}
Hence all matter-generated UV-divergences will be completely absorbed
in the gravitational counterterm coming from the action of WG. This
means that QWG allows for the quantum coupling of conformal matter to gravity
in a way preserving conformal symmetry. The main message is that the conformal
symmetry of classical matter models is not destroyed by quantum corrections
when coupled to conformal gravity. Moreover, gravity
enhances this coupling providing the whole uniform treatment of the
classical system, which is described in a fully conformally invariant
way.

One can introduce QWG completely from  the scratch along similar lines
of thoughts. For this, one needs to consider the path integral on curved fixed spacetime backgrounds, and  to integrate out the quantum fields
of conformal matter models (like electromagnetism, YM or SM Dirac
fermions before the electroweak symmetry breaking). As proved by 't Hooft and Veltman ~\cite{tHooft:1974toh, Hooft2}, this generates
the very specific action for quantum gravitational
theories. Since the conformal symmetry was originally in the matter
model, then the careful quantization procedure cannot break it and
it must be inherited by the resulting induced gravitational theory.
This must be the theory described by the $C^{2}$ action. Hence QWG
is inevitable, if one uses conformal matter models and if the conformal
symmetry is treated with sufficient care. If one did not know that
dynamical gravity is out there (like one did not know about the dynamical
theory of the gluonic field related to non-Abelian gauge symmetries),
then one would discover that theory (dynamical gravity) by integrating
out example massless Dirac fermions. And this dynamical quantum gravity
must be described by QWG theory. There is also a related line of argumentation
on how to rederive Einsteinian theory starting from Weyl gravity in
$d=4$ (due to Maldacena and others, see e.g.  \cite{Maldacena:2011mk}), but we will not discuss this
direction here, since it goes against the main spirit of our presentation,
although we admit that there are striking similarities between the
two theories (apparent for example on the level of exact vacuum solutions
discussed at length above). It is also interesting with this line of thought to study processes of black hole evaporation
within classical conformal gravity as it was done in \cite{Bambi:2017ott}.

Having discussed motivation for Weyl gravity as the theory of quantum
gravity, now we analyze properties of QWG understood as the QFT of gravitational
interactions with conformal symmetry. This last element is additional
and requires a special care. This is because the conformal symmetry
is in the local version here. It is not a global symmetry as it was
for example in highly (super)-symmetric ${\cal N}=4$ matter models.
Since it is in the gauged version, then this additional symmetry of
local character must be properly taken care of during the formal process
of quantization of gravity. The framework of Faddeev-Popov quantization
allows for this since it parallels the treatment of diffeomorphism
symmetries, which are from their nature local transformations. On
the other side, the conformal symmetry is here gauged (localized)
and there are some subtle differences compared to the other case of gauge
symmetries of relativistic gravitation (understood as the local gauge
theory of the Poincar\'e{} group). We refer the reader to a special
treatment of local gauge symmetries in this case. Of course, on the
quantum level one must secure that these local symmetries are not
anomalous. Otherwise, the whole program of quantization would miserably
fail. The issue of conformal anomaly is very important and we discuss
some comments about it at the end of this contribution.

For covariant quantization the best theoretical framework is that
one with functional path integrals. Since in gauge theory we have
a local symmetry and also degeneracy of the kinetic operator on the
level of path integral, then this symmetry must be constrained to
define properly the functional integral. One typically does a gauge
fixing. Here this should be done separately for diffeomorphisms and
for conformal symmetry of WG. The framework lets all this to be done
in a covariant manner. Next, one has to add appropriate Faddeev-Popov
quantum ghost fields. We remark that these ghost fields are good and
useful and they have nothing in common with the Boulware-Deser ghosts
of Weyl gravity (bad and malevolent ghosts). They have to be specific
to the diffeomorphism part of the symmetry group, as well as to the
conformal group. With such conformal ghosts new Feynman rules need
to be added to the perturbative rules of the quantum gravitational
theory turning it therefore into QWG theory. Moreover, finally we
can unambiguously find the propagator of the QWG graviton field when
some gauge choice is made. Of course, this propagator will depend
on the gauge fixing parameters used to kill the additional infinite
gauge freedom on the level of path integral.

The properties of QWG on the perturbative level of analysis in QFT
are very interesting. This quantum system gives rise to a renormalizable
model of quantum gravitation. The perturbative UV-divergences are
fully under control and they are all absorbable in the original covariant
terms of the theory, namely the known and seen before counterterm
$\sqrt{|g|}C^{2}$ of the original action of the theory. QWG has a
good control not only over divergences generated from the matter side,
it also constrains the divergences generated by quantum graviton loops.
The counterterm $R^{2}$ is not generated (at least on the level of
first quantum loop). The renormalizability of this model of quantum
conformal gravity is a very remarkable feature, and it persists even
after coupling to conformal matter models (which in $d=4$ are characterized
by dimensionless coupling constants). The QWG in its pure gravitational
sector has one unique coupling constant $\alpha_{C}$. Of course,
this coupling exhibits RG running due to the fact that the corresponding
$\beta$-function as read from the UV-divergences of the theory is
non-vanishing. However, asymptotically in the UV the control over
this running of $\alpha_{C}$ is gained since the QWG is asymptotically
free in deep UV regime. In this way, this model resembles very much
the situation met in QCD, where the theory also contains only one
positive unique coupling which in the UV limit is asymptotically free.

One knows that QCD as a YM theory of the non-Abelian group of $SU(3)$
is defined without any mass scale and the QCD coupling parameter is
dimensionless although it runs under RG. However, in QCD one sees
in the IR region spontaneous creation of mass scales, in the form
of generation of $\Lambda_{{\rm QCD}}$ or effectively the pion mass
$m_{\pi}$ (this last particle mediates effectively the strong interactions
between protons and neutrons at very low energies). This effect
is entirely due to the phenomenon known as dimensional transmutation.
Similar effects will take lead in QWG, where in the low energy
regime we should see a spontaneous generation of mass scales (like
the Planck scale), which describe effectively quantum gravitational
interactions at sufficiently low energies.

The fact that the QWG is asymptotically free in the ultraviolet regime
can be also seen as a particular case of a special situation when
the fixed point (FP) of RG is met in the UV. And the theory in the UV reaches some
CFT describing conformal gravitational interactions also on the quantum
level. The quest for such CFT theory with local conformal symmetry
and with gravitational interactions is very restricting. Especially,
the constraint that in the UV gravitational CFT there is no conformal
anomaly fixes the final UV theory quite uniquely. The theory in the
UV has to be an ${\cal N}=4$ conformal supergravity as discovered
by Fradkin and Tseytlin with four copies of ${\cal N}=4$ super-Yang-Mills
theories~\cite{Fradkin6}. In such a coupled model, there are no UV-divergences and
the quantum model is completely UV-finite. This framework of supergravity
is very constrained and the total symmetry algebra has many generators,
hence the divergences are constrained to vanish. Moreover, the conformal
current, the energy-momentum tensor and all other supersymmetric currents
find themselves in the same supersymmetry multiplet and that is why
they have to be simultaneously conserved, which implies the absence of
any violation of conformal symmetry, the preservation of the quantum structure
of CFT, the independence of any mass scale and UV-finiteness. Only with
extended supersymmetries it is possible to relate the conservation
of the energy-momentum tensor $\nabla_{\mu}T^{\mu\nu}=0$ to the conservation
of conformal current $\nabla_{\mu}j^{\mu}=0$ and hence to the situation
with unbroken scale and conformal invariance. In this very constrained
framework, the relativistic dynamics is very special, and for example
there is no dynamics related to RG flows. Still constraints put
a big amount of control on possible places from which divergences
could pop out requiring regularization. In such a constrained theory
it is still possible to perform a very well defined quantum computation,
which is the main result of this paper. For this it is enough to use
powers of the relativistic symmetries and of conformal symmetries
in the QFT theoretical framework.

The model due to Fradkin and Tseytlin realized fully the idea of quantum
conformality with gravitational interactions. In this way the conformality
of ${\cal N}=4$ SYM theory is extended away from flat spacetime to
a general curved background. It is very reassuring that when ${\cal N}=4$
SYM theory is put on curved background then even the conformally invariant
gravitational counterterm $\sqrt{|g|}C^{2}$ is not generated (of
course in the gauge sector the $F^{2}$ counterterm is not generated
to the flat spacetime UV-finiteness of this gauge theory model). Only
the coupling of the two theories (supergravity and super-Yang-Mills)
provides at the end the theory with no conformal (gauged) anomaly
on the quantum level. This theory is with quantum scale invariance
promoted to quantum conformal invariance. There are no beta functions,
no running couplings, no RG flows, only anomalous dimensions of quantum
operators. The conformal Ward identities are clearly preserved and
they constrain highly the Green functions. And as we discussed this
above, this theory sits all the time at the UV FP of RG, however the
breaking of conformal symmetry by adding some deformation operator
can be achieved. This will result with some non-trivial RG flows.
For some very particular situation, one can view this as a spontaneous
conformal symmetry breaking (induced and related to supersymmetry
breaking as well) and one can hope that when this symmetry is broken
in a soft way, then there is a remnant of some information in conformal
Ward identities of the theory even away from the UV FP, when the theory
was described by a gravitational CFT.

Another virtue of QWG in Euclidean framework is that the Euclidean
theory defined by action functional with $C^{2}$ as Lagrangian density
is bounded from below. This implies that the partition function of
the corresponding statistical mechanical model is completely positive-define
and well-defined, does not need any regularization, and the functional
integrals are mathematically very well-defined. Contrary to Euclidean
Einsteinian gravitational theory in QWG there is no conformal instability
problem, of course because here we do not have a conformal mode with
negative sign. In QWG we have only propagating modes which are invariant
under symmetry properties of the theory, so the conformal mode, (trace
of the metric tensor fluctuations) does not show up in the perturbative
spectrum. One can analyze with all available methods from statistical
mechanics and field theory the quantum partition function $Z$ here,
which is positive-definite and describes the quantum statistical partition
function. It is one of the very first consistent model of field theory
which in the context of quantum statistical mechanics gives statistical
and stochastic properties of Euclidean field theory. Especially, since
in Euclidean QWG this is a field theory with propagating metric degrees
of freedom describing differential geometry of $4$-dimensional manifolds
(surfaces). It could be said as an analogue of Euclidean gravity -
quantum statistical mechanical theory of differential manifolds. Since
the metric structure does not matter here, only the conformal properties,
and angles as mentioned in the introductions, this is really a study
of (scaleless) shape of manifolds. This theory can be viewed as a
first consistent model of quantum shape dynamics.

As emphasized a lot previously QWG is a system with a very peculiar
relativistic dynamics. Diffeomorphism symmetry is fully realized (symmetry
under the change of coordinate system used to describe physics). Moreover,
it is also conformal symmetry that it is gauged here (appears in the
local version). And this is a symmetry under an arbitrary changes
of the conformal factor used to measure distances. This puts a lot
of constraints on the ensuing relativistic dynamics of the gravitational
field. One could say from the historical perspective, that this is
a first modified gravitational theory (introduced by Weyl in 1918),
although it was not called like this neither by Weyl, nor by Einstein
- his opponent. His theory was clearly different than Einsteinian
gravity (introduced in 1918) because it embodies more symmetries and
the resulting dynamics is with higher derivatives, particularly in
$d=4$ this is with four derivatives, while Einstein theory gives
a dynamics with two derivatives only. One can see this difference
also on the level of the spectra of two theories. For example, in
QWG as analyzed on the tree-level the physical degrees of freedom
around any background are only transverse traceless (TT) gravitons
as this is obvious from the York decomposition (to be explained in
detail below). And we do not have for example a problematic conformal
mode of Einsteinian gravity.

Moreover, QWG is different than any other higher derivative (HD) gravitational
theory containing four derivatives in $d=4$ spacetime dimensions.
Such a general Stelle theory is characterized by two coupling parameters
and two covariant terms in the gravitational Lagrangian, namely $\alpha_{R}R^{2}+\alpha_{C}C^{2}$.
However, QWG is different and not a similar to a general quadratic
Stelle theory. The spectra are different since only in QWG we have
only TT gravitons present. One cannot also take in a simple continuous
way the limit $\alpha_{R}$ in the results of the general Stelle theory.
This is another incarnation of the famous Veltman discontinuities.
This time it is to be analyzed in HD gravitational theories. It is
not like original problem with the mass gap of gauge theories and
counting of propagating degrees of freedom there ($2$ in massless
vs $3$ in massive gauge theories). Here the discontinuity has to
do not only with transition from massive to massless excitations,
but also with a drastic change of the character of fluctuations in
the two theories. For example, they fill different representations
of the Lorentz group which is used to classify particle excitations
around flat spacetime. So the particles are in different tensorial
representation. For example, in QWG on the tree-level we have 2 massless
tensors and a massless vector. This last particle is not seen in any
other spectrum of ordinary HD gravitational theories in $d=4$. This
vector could be viewed as a remnant and the signal of the remaining
(vectorial) conformal symmetry on the level of the action of the theory.
One should see that in general the Veltman discontinuity means that
in the discontinuous limiting theory there exists an enhancement of
local symmetries. The theory with $\alpha_{R}=0$ is very special
and highly symmetric, higher than any ordinary HD gravitational theories.
And this is of course, due to the presence of local conformal symmetry
there.

\section{Quantum computations}

We would like to present here an example of a quantum computation
done in the framework of QWG. This computation has to be done in the
framework of quantum field theory (QFT) of gravitational (and conformal)
interactions. For this we will use the formalism of Faddeev-Popov
quantization where special care should be exerted in
the treatment of local conformal symmetry on the quantum level. We do
not wish that this symmetry is broken by the mere quantization
process. This quantum computation (at the one-loop level) will be
performed within the background-independent formalism and for this
we will use the background field method (BFM). Moreover, as our background
we will choose a general Einstein spacetime (ES) background, characterized
by the condition for Ricci tensor \begin{equation}
R_{\mu\nu}=\Lambda g_{\mu\nu},\label{eq: esdef}\end{equation}
 where $\Lambda$ is an arbitrary value constant parameter, that is
we have $\Lambda={\rm const}$ ($\nabla_{\mu}\Lambda=0$). The values
and the form of the Weyl tensor $C_{\mu\nu\rho\sigma}$ can be arbitrary
on general ES and only in this way the generic ES differs from a maximally
symmetric spacetimes (MSS) characterized by the parameter $\Lambda$.
Selecting a general ES background we get in one stroke the result of the previous computations
on two different backgrounds: namely on Ricci-flat
background and MSS. They were considered in \cite{Pjizba2}.

We will perform the computation  considering only the
physical degrees of freedom which are present as dynamical degrees
of freedom in QWG. As it is well known due to diffeomorphism symmetry
and due to the conformal symmetry they are transverse and traceless
(TT) gravitons respectively. Of course, a quantum graviton is described
as a fluctuation of the metric tensor of the spacetime, so these fluctuations
can be considered as a symmetric rank-2 tensor.

Our computation is an RG-improved one-loop level computation of the
RG running of couplings in QWG. In this calculation we take into account
the special role of non-perturbative phenomena (and this is why this is
an RG-improvement) encoded in the threshold phenomena (predominantly)
related to mass thresholds of quantum massive modes. Additionally
we also study the impact of the graviton's anomalous dimension on
these results. These are two non-perturbative effects that improve
over the standard quantum one-loop computation which was achieved in 1980's.
In this sense, we can say that our computation is a $1.5$-loop one
since we still acknowledge that the effects of two perturbative loops
will give a more accurate description of quantum phenomena. We start
our computation with the one-loop partition function of QWG on ES
and make it more sensitive to quantum effects by working with it in
the non-perturbative Wetterich equation describing the exact RG flows.
This means that our one-loop beta functions are functionally improved
in the sense of using functional Renormalization Group (FRG) approach.

The novelty of this approach is that for the first time we analyze
the general situation on ES, and not separately on MSS \cite{Irakleidou:2015exd} or Ricci-flat
backgrounds. This approach follows \cite{Benedetti:2009gn}. After analyzing the RG flow mainly in the deep IR regime
we will search for  FP's of RG (defined by condition that
beta functions vanish $\beta=0$). Finally, we discuss the physical
applications and interpretations of these results about the IR behavior
of QWG. This conclusion will open a new window for cosmology.

Before we embark on showing details of the computation, we should
explain why it is sensible to perform such quantum computation with
running couplings and beta function in QWG. We emphasized before that
if the conformality is preserved on the quantum level, then the theory
must be at a FP of RG, and there are no beta functions and no RG
flow. For consistency such a regime of the theory must be also conformal
anomaly-free. This implies that this theory has to be (very probably)
uniquely the superconformal gravity constructed by Fradkin and Tseytlin~\cite{Fradkin6}.
We believe such theory describes the conformal gravitational physics
in the deep UV regime, where all the symmetries (including supersymmetry
in local version) are restored. Therefore we find more interesting
to look at the deep UV behavior of the QWG where we can find some
non-trivial FP of RG describing probably new infrared physics. What
happens in the IR regime is the soft breaking of conformal symmetry
so that only the global part of the full $15$-dimensional conformal
group is broken. In particular, the generator related to global scale-invariance,
i.e. to the dilatation transformations on the physical system, is broken. This
breaking in a spontaneous way is a deformation of the original UV
conformal theory by giving a vev to some operator, which is globally
scale-invariant but not invariant under full conformal group. 

This is what may happen in QWG described by pure gravitational
action (without local supersymmetry) when analyzed consequently at
one-loop and two-loop perturbative levels of the theory. In the UV
this theory meets a UV FP of RG, where the scale-invariance is regained.
Also conformal anomaly vanishes there (due to participation of other
fields in the supermultiplet). When one lowers the energy scale and
decouples the superpartners, one sees that the quantum operator $R^{2}$
acquires a non-vanishing vev due to contribution at the two-loop order.
This is also related to the fact that the beta function $\beta_{R^{2}}$
of the coupling in front of the $R^{2}$ term appears on the second
perturbative loop order, while it was not present there at the first
loop order. In such a theory we start a non-trivial RG running, which
when approaching the IR limit, generates mass scales (like the Planck
scale) and the symmetries of the theory are changed. Likewise,
symmetries in QCD are changed when we start describing strong interactions
via interactions with massive pions.) This  happens because of the effect
of dimensional transmutation occurring in this theory. The quantum
theory which describes these phenomena towards IR regime of the spectrum
is QWG with non-vanishing beta functions of couplings and running
coupling parameters. Hence, it is useful to describe quantum RG effects
in such theory described by the quantized version of Weyl gravity,
when we still had the conformal symmetry on the classical level, and
when the quantization procedure did not break this and was done very
delicately. One sees that if sufficient care is exerted then conformal
Ward identities are still satisfied like in softly spontaneously broken
local gauge theories.

The issue of conformal anomaly (CA) does not prevent sensible computations
in such a theory since the computation is done towards the IR regime.
One notices that the CA is really an UV problem or in a different
disguise a problem with the UV-completion of the theory.
In other words, for a good definition of the theory one must find a UV FP with good
conformal properties. We already have discussed such  a candidate
theory, namely the ${\cal N}=4$ conformal supergravity (CSG). Moreover,
the CA at one-loop, for example, is completely expressed via the perturbative
beta functions of the theory. And it is well known that these beta
functions are related to the UV-divergences of the theory, so to the
UV limits of the theory. Hence physics in the IR does not matter for
this. If we assume that the UV-completion is done by CSG, then the
problem with CA is gone since this last one probes the UV physics.
We know that in IR the conformal symmetry has to be broken anyway,
but this will not influence at all the securement of CA issue in the
UV regime. With this understanding in mind, we can still reasonably
well and logically correct describe the situation with running in
the IR. In order to have the power of the remnants of the conformal
symmetry expressed on the quantum level via conformal Ward identities,
the conformal symmetry towards the IR regime has to be broken but
only in a softly way as described schematically above. Of course,
this violation of conformality has to be analyzed better and in more
details, and we leave it for future investigations. In this way, we
can still sensibly talk about RG flow in QWG in the IR regime, where
the softly violated conformal Ward identities still constrain the
physics in a sense that the quantized theory which takes into account
these phenomena in the IR regime is the quantized version of Weyl
gravity, and for example not a generic higher derivative (quadratic
in curvatures) gravitational theory. This is the perspective which
gives sense to the computation that we present below. This is exactly
a quantum computation in the IR regime of the QWG theory (non-supersymmetric).

In order to proceed with the quantum computation we need a few technical
details. First in order to get rid of ambiguities related to gauge
degrees of freedom and other non-physical degrees of freedom in QWG
and also explicit addition of Faddeev-Popov diffeomorphic and conformal
ghosts, we use the York decomposition of the metric fluctuations.
Working in BFM, we start with writing a general expansion of the full
quantum metric $\tilde{g}_{\mu\nu}$\begin{equation}
\tilde{g}_{\mu\nu}=g_{\mu\nu}+h_{\mu\nu},\end{equation}
where the fluctuations fields $h_{\mu\nu}$ are general and the background
metric $g_{\mu\nu}$ is a general ES metric. We perform the York decomposition
of the gravitational fluctuations in two steps. First, we remove the
spurious trace degree of freedom by writing that\begin{equation}
h_{\mu\nu}=\bar{h}_{\mu\nu}+\frac{1}{4}g_{\mu\nu}\phi\quad{\rm with}\quad\phi=h_{\mu}^{\mu}=g^{\mu\nu}h_{\mu\nu}\end{equation}
and with the tracelessness condition on the new fluctuations \begin{equation}
g^{\mu\nu}\bar{h}_{\mu\nu}=\bar{h}_{\mu}^{\mu}=0.\end{equation}
In the second step of York decomposition we remove the transverse
part from the traceless metric fluctuations $\bar{h}_{\mu\nu}$. This
is achieved by recalling the following formula\begin{equation}
\bar{h}_{\mu\nu}=\bar{h}_{\mu\nu}^{\perp}+\nabla_{\mu}\eta_{\nu}^{\perp}+\nabla_{\nu}\eta_{\mu}^{\perp}+\nabla_{\mu}\nabla_{\nu}\sigma-\frac{1}{4}g_{\mu\nu}\square\sigma,\end{equation}
where the transversality conditions on new metric fluctuations $\bar{h}_{\mu\nu}^{\perp}$
and on the vector field $\eta_{\mu}^{\perp}$ read\begin{equation}
\nabla^{\mu}\bar{h}_{\mu\nu}^{\perp}=0\quad{\rm and}\quad\nabla^{\mu}\eta_{\mu}^{\perp}=0.\label{eq: transversality}\end{equation}
Moreover, still the metric fluctuation field $\bar{h}_{\mu\nu}^{\perp}=h_{\mu\nu}^{TT}$
retains its traceless character\begin{equation}
g^{\mu\nu}\bar{h}_{\mu\nu}^{\perp}=\bar{h}_{\mu}^{\perp\mu}=0\label{eq: tracelessness}\end{equation}
and this implies that for convenience we can also redefine the trace
scalar field $\phi\to\bar{\phi}=\phi-\square\sigma$, where $\sigma$
is another scalar field. We notice that the final York decomposition
reads\begin{equation}
h_{\mu\nu}=\bar{h}_{\mu\nu}^{\perp}+\nabla_{\mu}\eta_{\nu}^{\perp}+\nabla_{\nu}\eta_{\mu}^{\perp}+\nabla_{\mu}\nabla_{\nu}\sigma-\frac{1}{4}g_{\mu\nu}\bar{\phi}.\label{eq: yorkdec}\end{equation}
The purpose of this decomposition is its main result namely the irreducible
part of the fluctuation, which is in the traceless transverse tensor
$\bar{h}_{\mu\nu}^{\perp}=h_{\mu\nu}^{TT}$. One can see that in the
QWG the physical field which propagates around any background metric
$g_{\mu\nu}$ (it does not even have to be an ES metric) is only a
traceless transverse graviton. This is a physical field of QWG, where
all redundancies due to the local gauge symmetries of the formalism
are removed. This feature is due to diffeomorphism (transversality)
and due to conformality (tracelessness). One can see it explicitly
when writing the second variation of the Weyl action functional (\ref{eq: weylgravaction})
around any on-shell background (a Bach-flat background). In such an on-shell situation the spurious quantum fields
$\eta_{\mu}^{\perp}$, $\sigma$ and $\bar{\phi}$ completely decouple
and they do not show up in the expression for the second variation
of the action integral. On a completely general background there appear
terms with fields $\eta_{\mu}^{\perp}$, $\sigma$ and $\bar{\phi}$
but they are all proportional to the Bach tensor $B^{\mu\nu}$, so
they correctly vanish when EOM in (\ref{eq: bacheom}) are used. But
below we consider only on-shell backgrounds (all ES are automatically
Bach-flat in $d=4$ spacetime dimensions).

Next, we discuss the most general expansion of terms quadratic in
generalized curvatures in $d=4$ dimensions. We consider here only
local scalar invariants of dimension four. This implies that in the
action they come only with dimensionless coupling coefficients, which
is important for the property of soft breaking of conformal symmetry.
We also analyze their properties under global and local scale transformations
and also their first variations.

The first invariant is known already to us conformal scalar invariant
$C^{2}$ whose expansion in other terms in curvature reads \begin{equation}
C^{2}=R_{\mu\nu\rho\sigma}^{2}-2R_{\mu\nu}^{2}+\frac{1}{3}R^{2}=2R_{\mu\nu}^{2}-\frac{2}{3}R^{2},\end{equation}
where in the last line we used the integration by parts under spacetime
volume integral and the Gauss-Bonnet topological relation. As we discussed
at length in the introduction, this invariant when properly densitized
is perfectly conformally invariant, which we express as the following
statement below. Its infinitesimal local conformal variation vanishes
identically, that is we have\begin{equation}
\delta_{c}\left(\sqrt{|g|}C^{2}\right)=0.\end{equation}
Of course, this is also true for finite conformal transformation which
is also a basis for CWG.

The next curvature scalar invariant that we can consider here is a
Gauss-Bonnet invariant (also known as topological Euler invariant).
Its expansion reads explicitly \begin{equation}
{\rm GB}=E=R_{\mu\nu\rho\sigma}^{2}-4R_{\mu\nu}^{2}+R^{2}.\label{eq: GBterm}\end{equation}
In four dimensions one can call it topological since it is related
by Gauss-Bonnet theorem to the topological (completely metric-independent)
characteristics of the spacetime, known as Euler topological invariant
or Euler characteristics $\chi_{4}$. This also implies that any its
variation under the integral integrates to zero, that is we have\begin{equation}
\delta\left(\int d^{4}x\sqrt{|g|}{\rm GB}\right)=0.\label{eq: GBvar}\end{equation}
In particular, this means that the variation is zero also under conformal
transformations, both infinitesimal and finite, local and global.

The last invariant we can call as a Starobinsky invariant since first
was invented by him in 1980 ~\cite{Starobinsky:1980te}. It is simply the square of the Ricci
curvature scalar $R^{2}$. Besides it being globally scale-invariant
(in $d=4$) related to the fact that the coefficient $\alpha_{R}$
is dimensionless, it also shows some more extended form of global
conformal invariance. Namely, it is invariant with respect to restricted
local conformal invariance. The general local conformal transformations
are parametrized by arbitrary scalar field $\Omega=\Omega(x)$ as
in (\ref{eq: conftrans}). However, the Starobinsky invariant $R^{2}$
does not change under conformal transformations if the parameter $\Omega$
satisfies the GR-covariant d'Alembertian (wave) equation, namely if
\begin{equation}
\square\Omega(x)=0.\end{equation}
This is less than general local conformal transformations with arbitrary
$\Omega$, but still it is more than just global scale-invariance
or no invariance at all. Therefore usage of this operator for the
breaking of the conformal symmetry in $d=4$ dimensions in the softest
possible way may be a preferable solution to get a soft RG flow in
QWG.

Now, we characterize the background ES manifolds. They are more general
than MSS and Ricci-flat spacetimes, but they include both these cases
as special subsets. To describe a general ES, one needs to specify
the value of the $\Lambda$ parameter in (\ref{eq: esdef}) and also
the form of the Weyl tensor $C_{\mu\nu\rho\sigma}$ consistent with
all symmetries (so in results 10 algebraically independent components
in $d=4$). As a consequence of defining condition of ES, we derive
the following curvature relations\begin{equation}
R=4\Lambda,\quad{\rm GB}=R_{\mu\nu\rho\sigma}^{2}\end{equation}
and since the expression for the Riemann tensor on a general ES is
given by \begin{equation}
R_{\mu\nu\rho\sigma}=\frac{\Lambda}{3}\left(g_{\mu\rho}g_{\nu\sigma}-g_{\mu\sigma}g_{\nu\rho}\right)+C_{\mu\nu\rho\sigma},\label{eq: riemes}\end{equation}
which is up to the Weyl tensor the same like on a general MSS with
$\Lambda$ parameter, we derive also relations for various curvature
square invariants\begin{equation}
C^{2}=R_{\mu\nu\rho\sigma}^{2}-\frac{8}{3}\Lambda^{2}\quad{\rm and}\quad{\rm GB}=C^{2}+\frac{8}{3}\Lambda^{2}.\end{equation}
We remark that in $d=4$ all ES are vacuum solutions to CWG, i.e.
they satisfy Bach EOM (\ref{eq: bacheom}). This is an important consequence
for the evaluation of the partition function  on on-shell backgrounds
and also for the possibility to use WKB approximation to describe
one-loop quantum physics. It has also implications for the form of
the second variation on the considered here backgrounds. In what follows
we will only consider classical on-shell backgrounds. However, quantum
perturbations do not have to satisfy classical EOM of the theory since
QFT effects are typically off-shell.

The next step consists of writing the expression for the one-loop
partition function in QWG on ES in terms of functional determinants
of some differential operators considered on this background. We first
start with writing the second variation of the conformal action of
classical conformal Weyl gravity. It reads\begin{equation}
\delta^{2}S_{{\rm conf}}=\!\int\! d^{4}x\sqrt{|g|}h^{TT\mu\nu}\left(\hat{\square}-\frac{2}{3}\Lambda\hat{1}+2\hat{C}\right)\left(\hat{\square}-\frac{4}{3}\Lambda\hat{1}+2\hat{C}\right)h_{\mu\nu}^{TT},\label{eq: secvar}\end{equation}
where we used York decomposition of fluctuations (\ref{eq: yorkdec})
and we commuted and integrated by parts covariant derivatives under
the volume spacetime integral. We also exploited the definition of
the Weyl tensor as a matrix $\hat{C}$ whose action on the graviton
fluctuations is the following\begin{equation}
\left(\hat{C}h\right)_{\mu\nu}=C_{\mu\alpha\nu\beta}h^{\alpha\beta}.\end{equation}
Finally, we define the identity matrix $\hat{1}$ in the space of
rank-2 symmetric tensor fluctuations by the formula\begin{equation}
\left(\hat{1}\right)^{\mu\nu}{}_{\alpha\beta}=\delta_{\alpha\beta}^{\mu\nu}=\delta_{(\alpha}^{\mu}\delta_{\beta)}^{\nu}=\frac{1}{2}\left(\delta_{\alpha}^{\mu}\delta_{\beta}^{\nu}+\delta_{\beta}^{\mu}\delta_{\alpha}^{\nu}\right).\end{equation}

The formula (\ref{eq: secvar}) was proven by an explicit computation
where we also used various non-trivial identities coming from taking
variational derivatives of the GB term in (\ref{eq: GBterm}). One
notices that for ES being on-shell background there is no participation
of spurious gauge-redundant fields of transverse vectors $\eta_{\mu}^{\perp}$
and two scalars $\sigma$ and $\bar{\phi}$. Moreover, one sees that
in the kernel of the expression in (\ref{eq: secvar}) between traceless
and transverse fluctuations $h_{\mu\nu}^{TT}$ there is a matrix-differential
operator, which has a direct factorization in a product of two single
two-derivative operators (shifted by some constant factors proportional
to $\Lambda$ and also by some matrix of the Weyl tensor $\hat{C}$.
It is remarkable that there is not a mixing term $\Lambda\hat{C}$
which could be allowed by dimensional reasons, and moreover such a
term separately would not be seen by taking a limit to MSS ($\hat{C}\to0$)
or to Ricci-flat spacetime ($\Lambda\to0$). This lets the factorization
of the kernel to be successful here. One can think additionally of
this kernel as the simplest generalization of the kernels that appear
separately in the expressions for the second variations on MSS and
on Ricci-flat spacetimes. In the last case, the second variation gets
simplified even to a quadratic form\begin{equation}
\delta^{2}S_{{\rm conf}}=\!\int\! d^{4}x\sqrt{|g|}h^{TT\mu\nu}\left(\hat{\square}+2\hat{C}\right)^{2}h_{\mu\nu}^{TT}.\end{equation}
One could naively derive the expression for the partition function
from the result in (\ref{eq: secvar}). The result would be \begin{equation}
\tilde{\tilde{Z}}_{{\rm 1-loop}}^{2}=\frac{1}{\det_{2TT}\left(\hat{\square}-\frac{2}{3}\Lambda\hat{1}+2\hat{C}\right)\det_{2TT}\left(\hat{\square}-\frac{4}{3}\Lambda\hat{1}+2\hat{C}\right)}.\end{equation}
But since in the theory we have local symmetries then the true expression
for the one-loop partition function is more complicated. We have to
take care of the Jacobian of the path integral variable transformation,
of fixing all the local gauge symmetries, and also of adding Faddeev-Popov
determinant (both for diffeomorphism and conformal part of the local
symmetry group here in QWG). At the end, one has to take care of possible
zero modes on ES and carefully exclude them from the computation of
determinants of the differential operators here.

When one uses only physical and gauge-invariant degrees of freedom
(only TT gravitons), then there is no need to additionally care about
fixing all the local gauge symmetries, and also about the addition of
Faddeev-Popov determinants. This is the advantage of using real degrees
of freedom which do not come with any redundancy related to local
symmetries. However, one still has to treat delicately the Jacobian
of the change of integration variables and the issue of zero modes.
The first problem of the functional change of the integration variables
under field theory path integral from original unconstrained spin-2
field $h_{\mu\nu}$ to a set of constrained fields $\left(h_{\mu\nu}^{TT},\eta_{\mu}^{\perp},\sigma,\bar{\phi}\right)$
is easily done even on a general spacetime background. For this issue,
one can easily understand that the situation on a general ES is the
same like on a general MSS with the same $\Lambda$ since the terms
in the Jacobian depend only on contractions of the Riemann tensor,
and not on the Riemann tensor itself. And one can understand from
the formula (\ref{eq: riemes}) that as for Ricci tensors and Ricci
scalars there is no any difference between MSS and ES. Therefore,
following what was done in the analysis of Jacobian on MSS we can
write\begin{equation}
\tilde{Z}_{{\rm 1-loop}}^{2}=\frac{\det_{1T}\left(\hat{\square}+\Lambda\hat{1}\right)\det_{0}\left(\hat{\square}+\frac{4}{3}\Lambda\hat{1}\right)}{\det_{2TT}\left(\hat{\square}-\frac{2}{3}\Lambda\hat{1}+2\hat{C}\right)\det_{2TT}\left(\hat{\square}-\frac{4}{3}\Lambda\hat{1}+2\hat{C}\right)}.\label{eq: partition1}\end{equation}
The above expression is closer to the final one for the one-loop partition
function on the general ES, but not yet the final one. We observe
that the factors in the numerator originate only from the analysis
on MSS background. However, they cannot be {}``corrected'' by the
addition of the matrix of the Weyl tensor $\hat{C}$ since this matrix
does not makes sense between spin-1 or spin-0 fluctuations, which
are used when we take determinants of the operators in the numerators.
These determinants in the subspaces of spin-1 and spin-0 fluctuations
respectively originate completely from the treatment of the Jacobian
of the field variables transformation (to the TT gravitational fluctuations).

The last problem is related to the presence of zero modes of the differential
operators which appear in (\ref{eq: partition1}) on a general ES
background. In full generality, such a problem is very complicated
to be tackled analytically. What we need actually, is a  much more modest
answer to the problem  of rewriting the determinants of operators
in the above formula. As it is well known zero modes appear mostly
due to constraints on operators that they have to act only in the
subspace of transverse fluctuations (both spin-1 and spin-2) since
these last constraints are of the differential character. The constraint
on the operator to act on traceless rank-2 tensor fluctuations $\bar{h}_{\mu\nu}$
does not create any problem for this since the last one is an algebraic
condition. Therefore it is desirable to change the determinants of
operators acting in subspaces of transverse fluctuations to determinants
of operators acting in extended subspaces, where the condition of
transversality is not imposed anymore. In our candidate expression
for the one-loop partition function on the ES background, we have
to change determinants in the space of spin-2 TT graviton fluctuations
and for spin-1 transverse vector fluctuations.

In general, formulas relating determinants in subspace of transverse
and extended fluctuations are derivable around any background. For
this purpose one must consider the following expressions\begin{equation}
\bar{h}^{\mu\nu}\square\bar{h}_{\mu\nu}\quad{\rm and}\quad\eta^{\mu}\square\eta_{\mu}\end{equation}
under the volume spacetime integrals and expand them using the York
decomposition for both traceless spin-2 fluctuations $\bar{h}_{\mu\nu}$
and for general unconstrained spin-1 fluctuations $\eta_{\mu}$ separately.
One has to commute derivatives, integrate them by parts, exploit various
tracelessness and transversality conditions as stipulated in formulas
(\ref{eq: transversality}) and (\ref{eq: tracelessness}). At the
end one has to use various curvature relations valid on a general
ES from (\ref{eq: esdef}) and (\ref{eq: riemes}).

One sees that the relation between determinants of operators for transverse
and unconstrained spin-1 fluctuations does depend only on the Ricci tensor
of the background spacetime. Hence in this place we again find no
difference between MSS and ES case. On MSS background spacetime, the
final formula reads\begin{equation}
\det{}_{1T}\left(\hat{\square}+X\hat{1}\right)=\frac{\det_{1}\left(\hat{\square}+X\hat{1}\right)}{\det_{0}\left(\hat{\square}+X\hat{1}-\Lambda\hat{1}\right)}\label{eq: zermod1}\end{equation}
and this formula allows to perform change from the subspace (1T) of
transverse spin-1 fluctuations to the space of unconstrained spin-1
fluctuations. The correction term is proportional to the functional
determinant of some scalar box operator, which receives also an additional
shift by the term proportional to the $\Lambda$ parameter (which
is the same for both ES and MSS backgrounds).

The situation with the sector of spin-2 TT fluctuations is more complicated.
The expressions for the expansion of $\bar{h}^{\mu\nu}\square\bar{h}_{\mu\nu}$
using the York decomposition on a generic ES background contains terms
proportional to the powers of the $\Lambda$ parameter as well as
terms proportional to the matrix of Weyl tensor $\hat{C}$. However,
the terms from the last group come without derivatives, they are only
of the endomorphic character without any remnant differential operator.
Again, one does not note any mixing terms (proportional for example
to $\Lambda\hat{C}$), which would be invisible in MSS or in Ricci-flat
limits. This signifies that the contribution to the exclusion of zero modes for differential operators is done separately from MSS backgrounds
(parametrized by the value of the $\Lambda$ parameter) and from Ricci-flat
backgrounds (parametrized by arbitrary Weyl tensor $\hat{C}$). Since
in the last case the resulting terms in $\bar{h}^{\mu\nu}\square\bar{h}_{\mu\nu}$
proportional to $\hat{C}$ are not of the differential characters,
then one concludes that there is no  contribution that has to be
subtracted from the determinants of the operators on Ricci-flat background
due to zero modes. The same happens on general ES and one can safely
forget about the contributions with Weyl tensor $\hat{C}$ there,
and only concentrate on the contribution that really have to be subtracted,
namely those which are proportional to the $\Lambda$ parameter of ES.
For this last aspect the computation is identical to the general case
of MSS. One ends up with the following formula\begin{equation}
\det{}_{2TT}\left(\hat{\square}+X\hat{1}\right)=\frac{\det_{2T}\left(\hat{\square}+X\hat{1}\right)}{\det{}_{1T}\left(\hat{\square}+X\hat{1}-\frac{5}{3}\Lambda\hat{1}\right)\det_{0}\left(\hat{\square}+X\hat{1}-\frac{8}{3}\Lambda\hat{1}\right)}\label{eq: zermod2}\end{equation}
and this formula allows to perform change from the subspace (2TT)
of traceless transverse spin-2 gravitational fluctuations to the space
of traceless spin-2 fluctuations. There are two correction terms.
They are proportional to the functional determinant of the differential
operator acting on transverse spin-1 fluctuations and to the functional
determinants of some scalar box operator, which receives also an additional
shift by the term proportional to the $\Lambda$ parameter on ES.
In accordance with the previous remarks, one sees in general that
such correction terms could not have a term with the matrix of the
Weyl tensor $\hat{C}$ because such matrix cannot act in the subspace
of spin-1 or spin-0 fluctuations due to simple algebraic reasons (there
is not a sufficient number of indices to contract with Weyl to produce
an endomorphism of the vector or scalar bundle).

Finally, one gets the expressions for the one-loop partition around
a general ES in the following form\begin{equation}
Z_{{\rm 1-loop}}^{2}=\frac{\det_{1}^{2}\left(\hat{\square}+\Lambda\hat{1}\right)\det_{1}\left(\hat{\square}+\frac{1}{3}\Lambda\hat{1}\right)\det_{0}\left(\hat{\square}+\frac{4}{3}\Lambda\hat{1}\right)}{\det_{2T}\left(\hat{\square}-\frac{2}{3}\Lambda\hat{1}+2\hat{C}\right)\det_{2T}\left(\hat{\square}-\frac{4}{3}\Lambda\hat{1}+2\hat{C}\right)\det_{0}\left(\hat{\square}+2\Lambda\hat{1}\right)}.\label{eq: partition2}\end{equation}
To get this expression we applied (\ref{eq: zermod1}) and (\ref{eq: zermod2})
to (\ref{eq: partition1}) and finally once again (\ref{eq: zermod1})
to the intermediate result. In this form, this expression resembles
very much the one obtained as the final one on MSS backgrounds. The
only difference is the addition of two terms with the matrix of the
Weyl tensor $\hat{C}$ in the denominator, when this algebraically
makes sense so only for spin-2 fluctuations. This is the final correct
expression for the one-loop partition function of QWG theory on ES
background. Its correctness was verified by performing the explicit
computation of the $b_{0}$, $b_{2}$ and $b_{4}$ coefficients related
to UV divergences of the QWG theory. For example, for the $b_{4}$
coefficient, they reproduce the famous results by Fradkin and Tseytlin
of divergences at one-loop level proportional to $\Lambda^{2}$ and
$C^{2}$ in one stroke.

One can study various limiting situations of the one-loop partition
function we have just computed in (\ref{eq: partition2}). First,
one can easily take the limit $\hat{C}\to0$ to reduce to MSS case.
This is only a small cosmetic modification since the terms with matrix
of the Weyl tensor $\hat{C}$ will be gone, but the structure of the
partition function $Z$ is exactly the same in MSS like in (\ref{eq: partition2}).
The limit $\Lambda\to0$ brings us back the situation on general Ricci-flat
spacetime. Then one sees that the form gets reduced to the following\begin{equation}
Z_{{\rm 1-loop},{\rm Ricci-flat}}^{2}=\frac{\det_{1}^{3}\left(\hat{\square}\right)}{\det_{2T}^{2}\left(\hat{\square}+2\hat{C}\right)}.\end{equation}
The last partition function can be extended to contain only fully
unconstrained fields of fluctuations (in particular of spin-2 fluctuations)
and then it reads\begin{equation}
Z_{{\rm 1-loop},{\rm Ricci-flat}}^{2}=\frac{\det_{1}^{3}\left(\hat{\square}\right)\det_{0}^{2}\left(\hat{\square}\right)}{\det_{2}^{2}\left(\hat{\square}+2\hat{C}\right)}.\end{equation}
The similar operation of extension can be also performed on the level
of the one-loop partition function on a general ES background. Namely,
we can write\begin{equation}
Z_{{\rm 1-loop}}^{2}=\frac{\det_{1}^{2}\left(\hat{\square}+\Lambda\hat{1}\right)\det_{1}\left(\hat{\square}+\frac{1}{3}\Lambda\hat{1}\right)\det_{0}\left(\hat{\square}-\frac{2}{3}\Lambda\hat{1}\right)\det_{0}\left(\hat{\square}-\frac{4}{3}\Lambda\hat{1}\right)\det_{0}\left(\hat{\square}+\frac{4}{3}\Lambda\hat{1}\right)}{\det_{2}\left(\hat{\square}-\frac{2}{3}\Lambda\hat{1}+2\hat{C}\right)\det_{2}\left(\hat{\square}-\frac{4}{3}\Lambda\hat{1}+2\hat{C}\right)\det_{0}\left(\hat{\square}+2\Lambda\hat{1}\right)}.\label{eq: partition3}\end{equation}
The two formulas in (\ref{eq: partition2}) and in (\ref{eq: partition3})
are completely equivalent due to the following extension formula\begin{equation}
\det{}_{2T}\left(\hat{\square}+X\hat{1}+2\hat{C}\right)=\frac{\det{}_{2}\left(\hat{\square}+X\hat{1}+2\hat{C}\right)}{\det{}_{0}\left(\hat{\square}+X\hat{1}\right)}.\end{equation}
This last formula has nothing to do with zero modes and this extension
of the operator to fully unconstrained spin-2 representation from
the traceless one. In this form this is valid on any spacetime. One
notices that due to algebraic reasons, the matrix of the Weyl tensor
$\hat{C}$ cannot appear in the denominator of the expression above.

The next step in the computation is the application of the functional
RG flow equation due to Wetterich~\cite{Wetterich:1993,Reuter:1998}. In a schematic form for a two-derivative
theory defined by the following kinetic part of the action in arbitrary
fluctuation field $\phi$ in whatever representation \begin{equation}
S_{{\rm kin}}=\int d^{4}x\sqrt{|g|}\phi\left(\hat{\square}+a\hat{1}\right)\phi\end{equation}
 the Wetterich equation takes the form\begin{equation}
k\partial_{k}\Gamma_{k}=\frac{1}{2}{\rm Tr}\left(\frac{k\partial_{k}R_{k}\hat{1}}{\hat{\square}+R_{k}\hat{1}+a\hat{1}}\right),\label{eq: wetterich}\end{equation}
where $\Gamma_{k}$ denotes the scale-dependent effective (average)
action for all fields in the theory, $a$ is a general constant shift
of the differential operator, and the cutoff kernel $R_{k}(z)$ is
a function of the leading part in derivatives of the differential
operator $z=\hat{\square}$ and also of the RG energy scale $k$.
The IR-cutoff kernel $R_{k}(z)$ leads to a suppression of the contribution
of modes with small eigenvalues of the covariant Laplacian operator
$\hat{\square}$ (namely these for which we have $-\hat{\square}\ll k^{2}$
in momentum space representation), while the factor $k\partial_{k}R_{k}\hat{1}$
in the numerator of the Wetterich equation removes contributions from
large eigenvalues $-\hat{\square}\gg k^{2}$. In this way, the loop
integrals to be done as functional traces in Wetterich equation are
both IR- and UV-finite. The functional traces in (\ref{eq: wetterich})
are traces both in the spacetime integration sense as well as in the
normal traces over all internal indices the fields may carry with
themselves. In our case, this will be traces over Lorentz indices
on metric fluctuations $h_{\mu\nu}$ (or on traceless fluctuations
$\bar{h}_{\mu\nu}$) and on vector field fluctuations. Obviously,
for scalars such internal traces are identical to multiplication by
unity.

Following the developments of the Wetterich equation for the constrained
fields and for the quantum theory read from the expression of its
one-loop partition function understood as product of various functional
determinants (raised to various powers) of simple two-derivative operators
$\hat{\square}$ possibly shifted by some constant endomorphic terms
$a\hat{1}$, the resulting reduced Wetterich equation looks as follows\begin{equation}
k\partial_{k}\Gamma_{k}=\frac{1}{2}\sum_{i}\sum_{j}\pm{\rm Tr}_{\phi_{i}}\left(\frac{\left(k\partial_{k}R_{k}-\eta R_{k}\right)\hat{1}}{\hat{\square}+R_{k}\hat{1}+a_{i,j}\hat{1}}\right)\!,\label{eq: wetterichred}\end{equation}
where we have denoted the anomalous dimension (identical for all fluctuation
fields) as $\eta$. Additionally, the $\pm$ signs depend on what
was in the exponent on the corresponding term in the one-loop partition
function (\ref{eq: partition2}) or (\ref{eq: partition3}). In other
words, this decides whether the general factor of the type $\left(\hat{\square}+R_{k}\hat{1}+a_{i,j}\hat{1}\right)$
was originally in the denominator or the numerator of the partition
function, respectively and with which power. The functional traces
were separately considered in each subspace of fluctuation fields
$\phi_{i}$, and in each subsector of fluctuations we allowed for
different factors of the type $\left(\hat{\square}+R_{k}\hat{1}+a_{i,j}\hat{1}\right)$
with different shifts $a_{i,j}$ (the index $j$ here is internal
to each subsector of fluctuations, while the index $i$ counts different
subspaces with different representation of fields $\phi_{i}$). The
correctness of the above steps that lead to the above FRG flow equation
in a reduced form (\ref{eq: wetterichred}) is, for example, verified
by an independent check of the coefficients for one-loop perturbative
beta functions of dimensionless coupling parameters in QWG.

One should write the Wetterich equation in a reduced form in an explicit
way on a general ES background using the two form of the one-loop
partition function. Here are the results of these expansions. From
formula in (\ref{eq: partition2}) one gets\[
k\partial_{k}\Gamma_{k}=\frac{1}{2}{\rm Tr}_{2T}\left(\frac{\left(k\partial_{k}R_{k}-\eta R_{k}\right)\hat{1}}{\hat{\square}-\frac{2}{3}\Lambda\hat{1}+2\hat{C}}\right)+\frac{1}{2}{\rm Tr}_{2T}\left(\frac{\left(k\partial_{k}R_{k}-\eta R_{k}\right)\hat{1}}{\hat{\square}-\frac{4}{3}\Lambda\hat{1}+2\hat{C}}\right)+\frac{1}{2}{\rm Tr}_{0}\left(\frac{\left(k\partial_{k}R_{k}-\eta R_{k}\right)\hat{1}}{\hat{\square}+2\Lambda\hat{1}}\right)-\]
\begin{equation}
-{\rm Tr}_{1}\left(\frac{\left(k\partial_{k}R_{k}-\eta R_{k}\right)\hat{1}}{\hat{\square}+\Lambda\hat{1}}\right)-\frac{1}{2}{\rm Tr}_{1}\left(\frac{\left(k\partial_{k}R_{k}-\eta R_{k}\right)\hat{1}}{\hat{\square}+\frac{1}{3}\Lambda\hat{1}}\right)-\frac{1}{2}{\rm Tr}_{0}\left(\frac{\left(k\partial_{k}R_{k}-\eta R_{k}\right)\hat{1}}{\hat{\square}+\frac{4}{3}\Lambda\hat{1}}\right),\label{eq: flow1}\end{equation}
while when we use the alternative (but equivalent) expression from
(\ref{eq: partition3}) one finds instead a slightly more complicated
formula which reads\[
k\partial_{k}\Gamma_{k}=\frac{1}{2}{\rm Tr}_{2T}\left(\frac{\left(k\partial_{k}R_{k}-\eta R_{k}\right)\hat{1}}{\hat{\square}-\frac{2}{3}\Lambda\hat{1}+2\hat{C}}\right)+\frac{1}{2}{\rm Tr}_{2T}\left(\frac{\left(k\partial_{k}R_{k}-\eta R_{k}\right)\hat{1}}{\hat{\square}-\frac{4}{3}\Lambda\hat{1}+2\hat{C}}\right)+\frac{1}{2}{\rm Tr}_{0}\left(\frac{\left(k\partial_{k}R_{k}-\eta R_{k}\right)\hat{1}}{\hat{\square}+2\Lambda\hat{1}}\right)-\]
\[
-{\rm Tr}_{1}\left(\frac{\left(k\partial_{k}R_{k}-\eta R_{k}\right)\hat{1}}{\hat{\square}+\Lambda\hat{1}}\right)-\frac{1}{2}{\rm Tr}_{1}\left(\frac{\left(k\partial_{k}R_{k}-\eta R_{k}\right)\hat{1}}{\hat{\square}+\frac{1}{3}\Lambda\hat{1}}\right)-\frac{1}{2}{\rm Tr}_{0}\left(\frac{\left(k\partial_{k}R_{k}-\eta R_{k}\right)\hat{1}}{\hat{\square}+\frac{4}{3}\Lambda\hat{1}}\right)-\]
\begin{equation}
-\frac{1}{2}{\rm Tr}_{0}\left(\frac{\left(k\partial_{k}R_{k}-\eta R_{k}\right)\hat{1}}{\hat{\square}-\frac{2}{3}\Lambda\hat{1}}\right)-\frac{1}{2}{\rm Tr}_{0}\left(\frac{\left(k\partial_{k}R_{k}-\eta R_{k}\right)\hat{1}}{\hat{\square}-\frac{4}{3}\Lambda\hat{1}}\right).\label{eq: flow2}\end{equation}
In the next step, we need to write and simplify the form of the beta
functional (LHS of the Wetterich equation) on a general ES background.
This is of course, related to the choice of the action ansatz for
the running effective action $\Gamma_{k}$ and to our abilities to
read various beta functions from it.

In general, we are interested only in running of dimensionless couplings
since their RG flow appears first when the conformal symmetry is broken
softly and spontaneously. In our $4$-dimensional setup such couplings
appear in front of three quadratic invariants in curvatures which
we defined above and which are consistent with diffeomorphism symmetry
of the quantum theory (since conformal symmetry is already softly
violated by the mere presence of RG flow for dimensionless couplings).
Therefore the general beta functional of the theory may have the form\begin{equation}
k\partial_{k}\Gamma_{k}=\!\int\! d^{4}x\sqrt{|g|}\left(\beta_{R}R^{2}+\beta_{C}C^{2}+\beta_{{\rm GB}}{\rm GB}\right).\end{equation}
In the above formula we defined that the general running effective
action had the form\begin{equation}
\Gamma_{k}=\!\int\! d^{4}x\sqrt{|g|}\left(\alpha_{R}(k)R^{2}+\alpha_{C}(k)C^{2}+\alpha_{{\rm GB}}(k){\rm GB}\right)\end{equation}
and that the beta functions of the running couplings parameter $\alpha_{i}=\alpha_{i}(k)$
are defined as logarithmic derivative of them with respect to RG scale
$k$ that is the following formula holds

\begin{equation}
\beta_{i}=k\frac{d}{dk}\alpha_{i}(k).\end{equation}
We notice that as it is common in some FRG approaches the $k$-dependence
for the one-loop RG-improved effective action $\Gamma_{k}$ sits only
in the running couplings, while the diffeomorphically invariant terms
there are not touched at all by the derivative with respect to RG
scale $k$ needed when defining non-perturbatively the beta function$\beta_{i}$.
Moreover, this is not the end of our choice of the general ansatz
on ES. It is well known that perturbatively in QWG the beta function
of the Starobinsky $R^{2}$ term is highly suppressed compared to
other beta functions $\beta_{C}$ and $\beta_{{\rm GB}}$ (that is
we have parametrically that $\beta_{R}\ll\beta_{C},\,\beta_{{\rm GB}}$).
This hierarchy of the system of beta functions is already seen on
the level of first loop. There holds that $\beta_{R}=0$ exactly to
one-loop level. If one considers higher loops then the expression
for $\beta_{R}$ is smaller because it contains higher powers of the
perturbative expansion parameter which for us here is assumed to be
small. These parameters are the couplings of the original theory.
Following this important observation we may neglect in our computation
the presence of the $\beta_{R}$ function both on the perturbative
and also on non-perturbative levels. Therefore our reduced beta functional
of the theory describing running effective action $\Gamma_{k}$ has
the form\begin{equation}
k\partial_{k}\Gamma_{k}=\!\int\! d^{4}x\sqrt{|g|}\left(\beta_{C}C^{2}+\beta_{{\rm GB}}{\rm GB}\right)\label{eq: betafnl}\end{equation}
 derived from\begin{equation}
\Gamma_{k}=\!\int\! d^{4}x\sqrt{|g|}\left(\alpha_{C}(k)C^{2}+\alpha_{{\rm GB}}(k){\rm GB}\right),\end{equation}
while we unanimously agree that on the RHS of the Wetterich flow equation
we have the partition function coming exactly from the QWG action
functional described by the truncation ansatz $\alpha_{C}C^{2}$ in
agreement with (\ref{eq: weylgravaction}). On the RHS of the FRG
flow equation the full $k$-dependence is brought about by IR cutoff
kernels $R_{k}$ and we decide not to include any additional $k$-dependence
in the Weyl coupling $\alpha_{C}$ on this RHS (a source side which
is a reason for all effects of RG running of the LHS of the RG flow
equation).

The beta functional (\ref{eq: betafnl}) evaluated on the generic
ES background gives\begin{equation}
\left.\beta_{C}C^{2}+\beta_{{\rm GB}}{\rm GB}\right|_{{\rm ES}}=\beta_{C}C^{2}+\beta_{{\rm GB}}\left(C^{2}+\frac{8}{3}\Lambda^{2}\right)=\left(\beta_{C}+\beta_{{\rm GB}}\right)C^{2}+\frac{8}{3}\beta_{{\rm GB}}\Lambda^{2}\label{eq: betafnlES}\end{equation}
Therefore on a general ES manifold we can read in principle independently
two beta functions $\beta_{C}$ and $\beta_{{\rm GB}}$. This is possible
due to an extraction of terms proportional to the invariant quantities
$C^{2}$ and $\Lambda^{2}$ respectively from the RHS of the FRG flow
equation in (\ref{eq: wetterich}). Again, we emphasize that the advantage
of using ES as background spaces is that we read these two beta functions
from one expression evaluated on one single background, hence there
is no any issue related to changing the background and to mutual interrelations
and interdependences between them. Of course, such a background should
be understood only as a theoretical tool and a trick which is very
helpful in extracting the beta functions of the quantum theory. One
should not think that our theory has to be considered always on ES
background. For consistency of the quantum theory the QWG in the form
we present here can be considered on general Bach-flat backgrounds.

In order to proceed with the quantum computation of beta functions
in the one-loop RG improved scheme due to non-perturbative FRG phenomena,
one has to make a choice for the explicit cutoff kernel function $R_{k}(z)$.
We follow the standard choice given by the Litim cutoff\begin{equation}
R_{k}=R_{k}(z)=\left(k^{2}-z\right)\theta\left(k^{2}-z\right),\end{equation}
where $\theta(x)$ denotes standard Heaviside theta step function.
For such a choice of the cutoff kernel we find that the functional
traces (the spacetime integration part of them) of some general function
of the general at most two-derivative operator $\hat{{\cal O}}$ is
given by\begin{equation}
{\rm Tr}f\left(\hat{{\cal O}}\right)=B_{4}\left(\hat{{\cal O}}\right)Q_{0}[f],\label{eq: trf}\end{equation}
where the $B_{4}$ coefficients of any differential operator are well
known in QFT and in the theory of operators. The $Q_{0}$ functional
of the function $f$ can be simplified in our case of interest, if
we recall what is the generic form of the function of the operator
$f=f(z)$ that we usually meet in our consideration related to the
Wetterich equation. This form reads\begin{equation}
f(z)=\frac{k\partial_{k}R_{k}(z)}{z+R_{k}(z)+a_{i,j}}\end{equation}
for a generic two-derivative factor of the type $\left(\hat{\square}+R_{k}\hat{1}+a_{i,j}\hat{1}\right)$.
In our case we also accept the identification that $\hat{{\cal O}}=\hat{\square}$
as the leading in the number of derivatives part of the operator $\hat{\square}+a_{i,j}\hat{1}$.
Then in such circumstances the expression for $Q_{0}$ functional
simplifies to\begin{equation}
Q_{0}[f]=Q_{0}\left[\frac{k\partial_{k}R_{k}(z)}{z+R_{k}(z)+a_{i,j}}\right]=(2-\eta)\left(1+\frac{a_{i,j}}{k^{2}}\right)^{-1}.\label{eq: q0fnl}\end{equation}
Using such a form of IR-decoupling we are sure that we correctly take
care of the threshold effects related to massive modes of the excitations
present around a given background. This decoupling is explicitly realized
by the second factor in the above formula, while the first factor
describes the corrections due to the inclusion of the anomalous dimension
for the quantum fields. Here $\eta$ denotes the uniform anomalous
dimension for the graviton field, since all TT graviton fields, vector
fields and scalars must have it the same (due to the requirements
of original conformal symmetry of the initial theory and due to always
unbroken diffeomorphism symmetry).

With the formulas (\ref{eq: trf}) and (\ref{eq: q0fnl}) at hand,
one can easily compute the ensuing functional traces in the FRG flow
equation. First, we write it for the compact form of the one-loop
partition function on a general ES given in (\ref{eq: flow1}). Then
the flow simplifies to\[
\frac{2}{2-\eta}k\partial_{k}\Gamma_{k}=\left(1-\frac{\frac{2}{3}\Lambda}{k^{2}}\right)^{-1}B_{4}\left(\hat{\square}_{2T}-\frac{2}{3}\Lambda\hat{1}_{2T}+2\hat{C}\right)+\left(1-\frac{\frac{4}{3}\Lambda}{k^{2}}\right)^{-1}B_{4}\left(\hat{\square}_{2T}-\frac{4}{3}\Lambda\hat{1}_{2T}+2\hat{C}\right)+\]
\[
+\left(1+\frac{2\Lambda}{k^{2}}\right)^{-1}B_{4}\left(\hat{\square}_{0}+2\Lambda\hat{1}_{0}\right)-\left(1+\frac{\frac{4}{3}\Lambda}{k^{2}}\right)^{-1}B_{4}\left(\hat{\square}_{0}+\frac{4}{3}\Lambda\hat{1}_{0}\right)-\]
\begin{equation}
-2\left(1+\frac{\Lambda}{k^{2}}\right)^{-1}B_{4}\left(\hat{\square}_{1}+\Lambda\hat{1}_{0}\right)-\left(1+\frac{\frac{1}{3}\Lambda}{k^{2}}\right)^{-1}B_{4}\left(\hat{\square}_{1}+\frac{1}{3}\Lambda\hat{1}_{0}\right).\label{eq: flowb41}\end{equation}
In the case of the alternative equation for the one-loop partition
function given in (\ref{eq: flow2}) we instead find\[
\frac{2}{2-\eta}k\partial_{k}\Gamma_{k}=\left(1-\frac{\frac{2}{3}\Lambda}{k^{2}}\right)^{-1}B_{4}\left(\hat{\square}_{2}-\frac{2}{3}\Lambda\hat{1}_{2}+2\hat{C}\right)+\left(1-\frac{\frac{4}{3}\Lambda}{k^{2}}\right)^{-1}B_{4}\left(\hat{\square}_{2}-\frac{4}{3}\Lambda\hat{1}_{2}+2\hat{C}\right)+\]
\[
+\left(1+\frac{2\Lambda}{k^{2}}\right)^{-1}B_{4}\left(\hat{\square}_{0}+2\Lambda\hat{1}_{0}\right)-\left(1+\frac{\frac{4}{3}\Lambda}{k^{2}}\right)^{-1}B_{4}\left(\hat{\square}_{0}+\frac{4}{3}\Lambda\hat{1}_{0}\right)-\]
\[
-\left(1-\frac{\frac{2}{3}\Lambda}{k^{2}}\right)^{-1}B_{4}\left(\hat{\square}_{0}-\frac{2}{3}\Lambda\hat{1}_{0}\right)-\left(1-\frac{\frac{4}{3}\Lambda}{k^{2}}\right)^{-1}B_{4}\left(\hat{\square}_{0}-\frac{4}{3}\Lambda\hat{1}_{0}\right)-\]
\begin{equation}
-2\left(1+\frac{\Lambda}{k^{2}}\right)^{-1}B_{4}\left(\hat{\square}_{1}+\Lambda\hat{1}_{0}\right)-\left(1+\frac{\frac{1}{3}\Lambda}{k^{2}}\right)^{-1}B_{4}\left(\hat{\square}_{1}+\frac{1}{3}\Lambda\hat{1}_{0}\right).\label{eq: flowb42}\end{equation}
However, already on this level one sees that these two forms of the
FRG flows are equivalent because of the following formula\begin{equation}
B_{4}\left(\hat{\square}_{2T}-X\hat{1}_{2T}+2\hat{C}\right)=B_{4}\left(\hat{\square}_{2}-X\hat{1}_{2}+2\hat{C}\right)-B_{4}\left(\hat{\square}_{0}-X\hat{1}_{0}\right)\end{equation}
for any constant $X$. Therefore for future considerations, we will
consider only the explicit RG flow given by the formula (\ref{eq: flowb41}).
In the formulas above one explicitly sees the presence of massive
threshold functions of the general type \begin{equation}
\left(1+\frac{a_{i,j}}{k^{2}}\right)^{-1}.\end{equation}
As we will see in a moment they are very important for the consideration
of the RG flows towards the deep IR regime in QWG. To eventually simplify
the expression for the RG flow and read the non-perturbative expressions
for the beta functions with the threshold phenomena included, one
needs to evaluate the $B_{4}$ coefficients related to perturbative
UV-divergences of the theory. Precisely they are related to logarithmic
dimensionless UV-divergences in front of the terms proportional to
$C^{2}$ and ${\rm GB}$ term in the quantum theory.

Explicit computations on the general ES give us the results for these
$B_{4}$ coefficients of various operators\begin{equation}
B_{4}\left(\hat{\square}_{2T}-\frac{2}{3}\Lambda\hat{1}_{2T}+2\hat{C}\right)=\frac{21}{20}C^{2}-\frac{7}{5}\Lambda^{2}\label{eq: b41}\end{equation}
\begin{equation}
B_{4}\left(\hat{\square}_{2T}-\frac{4}{3}\Lambda\hat{1}_{2T}+2\hat{C}\right)=\frac{21}{20}C^{2}+\frac{3}{5}\Lambda^{2}\end{equation}
\begin{equation}
B_{4}\left(\hat{\square}_{1}+\Lambda\hat{1}_{0}\right)=-\frac{11}{180}C^{2}+\frac{716}{135}\Lambda^{2}\end{equation}
\begin{equation}
B_{4}\left(\hat{\square}_{1}+\frac{1}{3}\Lambda\hat{1}_{0}\right)=-\frac{11}{180}C^{2}+\frac{236}{135}\Lambda^{2}\end{equation}
\begin{equation}
B_{4}\left(\hat{\square}_{0}+2\Lambda\hat{1}_{0}\right)=\frac{1}{180}C^{2}+\frac{479}{135}\Lambda^{2}\end{equation}
\begin{equation}
B_{4}\left(\hat{\square}_{0}+\frac{4}{3}\Lambda\hat{1}_{0}\right)=\frac{1}{180}C^{2}+\frac{269}{135}\Lambda^{2}\label{eq: b46}\end{equation}
We conveniently above wrote these formulas in terms of $C^{2}$ and
$\Lambda^{2}$invariant to be prepared for the extraction of two beta
functions $\beta_{C}$ and $\beta_{{\rm GB}}$. Once again one notices
that in these formulas there are no interference terms between $\Lambda$
and matrix $\hat{C}$ terms. Of course, they are not possible due
to algebraic reasons here, because the final expression for the $B_{4}$
coefficient must be a scalar and the matrix $\hat{C}$ is completely
traceless. On the other hand this result signifies that the contributions
from the MSS part and from the Ricci-flat part to the general ES are
independent of each other and there is no any mixing between them
and also that they can be computed separately.

Now, we read the beta functions exploiting together formulas (\ref{eq: flowb41}),
(\ref{eq: betafnl}), (\ref{eq: betafnlES}) and a set of formulas
from (\ref{eq: b41}) to (\ref{eq: b46}). We therefore find that
the following equation holds\[
\frac{2}{2-\eta}\left(\left(\beta_{C}+\beta_{{\rm GB}}\right)C^{2}+\frac{8}{3}\beta_{{\rm GB}}\Lambda^{2}\right)=\left(1-\frac{\frac{2}{3}\Lambda}{k^{2}}\right)^{-1}\left(\frac{21}{20}C^{2}-\frac{7}{5}\Lambda^{2}\right)+\left(1-\frac{\frac{4}{3}\Lambda}{k^{2}}\right)^{-1}\left(\frac{21}{20}C^{2}+\frac{3}{5}\Lambda^{2}\right)+\]
\[
+\left(1+\frac{2\Lambda}{k^{2}}\right)^{-1}\left(\frac{1}{180}C^{2}+\frac{479}{135}\Lambda^{2}\right)-\left(1+\frac{\frac{4}{3}\Lambda}{k^{2}}\right)^{-1}\left(\frac{1}{180}C^{2}+\frac{269}{135}\Lambda^{2}\right)-\]
\begin{equation}
-2\left(1+\frac{\Lambda}{k^{2}}\right)^{-1}\left(-\frac{11}{180}C^{2}+\frac{716}{135}\Lambda^{2}\right)-\left(1+\frac{\frac{1}{3}\Lambda}{k^{2}}\right)^{-1}\left(-\frac{11}{180}C^{2}+\frac{236}{135}\Lambda^{2}\right).\label{eq: flow5}\end{equation}
By comparing both sides of this equation and treating the two invariants
$C^{2}$ and $\Lambda^{2}$ as exclusive and mutually independent,
we arrive to the two equations\[
\frac{2}{2-\eta}\left(\frac{8}{3}\beta_{{\rm GB}}\Lambda^{2}\right)=\left(1-\frac{\frac{2}{3}\Lambda}{k^{2}}\right)^{-1}\left(-\frac{7}{5}\Lambda^{2}\right)+\left(1-\frac{\frac{4}{3}\Lambda}{k^{2}}\right)^{-1}\left(\frac{3}{5}\Lambda^{2}\right)+\]
\[
+\left(1+\frac{2\Lambda}{k^{2}}\right)^{-1}\left(\frac{479}{135}\Lambda^{2}\right)-\left(1+\frac{\frac{4}{3}\Lambda}{k^{2}}\right)^{-1}\left(\frac{269}{135}\Lambda^{2}\right)-\]
\begin{equation}
-2\left(1+\frac{\Lambda}{k^{2}}\right)^{-1}\left(\frac{716}{135}\Lambda^{2}\right)-\left(1+\frac{\frac{1}{3}\Lambda}{k^{2}}\right)^{-1}\left(\frac{236}{135}\Lambda^{2}\right)\label{eq: flowMSS}\end{equation}
 and\begin{equation}
\frac{2}{2-\eta}\left(\beta_{C}+\beta_{{\rm GB}}\right)C^{2}=2\times\left(\frac{21}{20}C^{2}\right)+\left(\frac{1}{180}C^{2}\right)-\left(\frac{1}{180}C^{2}\right)-3\times\left(-\frac{11}{180}C^{2}\right).\label{eq: flowrf}\end{equation}
These two equations represent respectively the RG flows as considered
separately on MSS and Ricci-flat backgrounds. In order to obtain the
equation in (\ref{eq: flowrf}) we also took a zeroth order series
coefficient in expansion in the $\Lambda$ parameter, since this parameter
is non-vanishing only on MSS. In this way we prove that the results
as obtained on a general ES completely reproduce the previous results
obtained separately on MSS and Ricci-flat backgrounds. This also confirms
that the original method with using of two distinct backgrounds is
consistent and does not lead to any contradiction. These two backgrounds
are mutually exclusive since the MSS can be Ricci-flat only if this
is a flat spacetime, however this observation regarding the physical
background does not lead to any inconsistency here. We used these
two backgrounds as a trick to get two combinations of beta functions
but none should understand that we consider a QWG theory put simultaneously
on these two backgrounds. The results we get there are consistent
with each other. Moreover, they are also consistent with the general
considerations that we make on ES, when the background used again
as a mathematical trick is a single one. However, we do not find any
problem in taking respective limits $\Lambda\to0$ or $\hat{C}\to0$
and we recover previous results from Ricci-flat and MSS respectively.
Moreover, there are no mixing terms $\Lambda\hat{C}$ which would
invalidate the independent analysis of MSS and Ricci-flat to get beta
functions $\beta_{C}$ and $\beta_{{\rm GB}}$ separately. This verifies
a general philosophy that the quantum beta functions of the theory
(in a given scheme) are universal and one can use arbitrary backgrounds
to evaluate them. Before we used MSS and Ricci-flat because of the
reasons of convenience. Now, we have proven that the same consistent
results are also obtained on the general ES.

By doing algebraic simplifications in the formula one finally arrives
to the system\[
\frac{2}{2-\eta}\beta_{{\rm GB}}\Lambda^{2}=\left(1-\frac{\frac{2}{3}\Lambda}{k^{2}}\right)^{-1}\left(-\frac{21}{40}\Lambda^{2}\right)+\left(1-\frac{\frac{4}{3}\Lambda}{k^{2}}\right)^{-1}\left(\frac{9}{40}\Lambda^{2}\right)+\left(1+\frac{2\Lambda}{k^{2}}\right)^{-1}\left(\frac{479}{360}\Lambda^{2}\right)-\]
\begin{equation}
-\left(1+\frac{\frac{4}{3}\Lambda}{k^{2}}\right)^{-1}\left(\frac{269}{360}\Lambda^{2}\right)-\left(1+\frac{\Lambda}{k^{2}}\right)^{-1}\left(\frac{179}{45}\Lambda^{2}\right)-\left(1+\frac{\frac{1}{3}\Lambda}{k^{2}}\right)^{-1}\left(\frac{59}{90}\Lambda^{2}\right)\label{eq: flowMSS2}\end{equation}
 and\begin{equation}
\frac{2}{2-\eta}\left(\beta_{C}+\beta_{{\rm GB}}\right)C^{2}=\frac{137}{60}C^{2}.\label{eq: flowrf2}\end{equation}
From this system one solves for the beta functions $\beta_{{\rm GB}}$
and $\beta_{C}$ in the following way\[
\beta_{{\rm GB}}=\frac{1}{2}(2-\eta)\left[-\frac{21}{40}\left(1-\frac{\frac{2}{3}\Lambda}{k^{2}}\right)^{-1}+\frac{9}{40}\left(1-\frac{\frac{4}{3}\Lambda}{k^{2}}\right)^{-1}+\frac{479}{360}\left(1+\frac{2\Lambda}{k^{2}}\right)^{-1}-\right.\]
\begin{equation}
\left.-\frac{269}{360}\left(1+\frac{\frac{4}{3}\Lambda}{k^{2}}\right)^{-1}-\frac{179}{45}\left(1+\frac{\Lambda}{k^{2}}\right)^{-1}-\frac{59}{90}\left(1+\frac{\frac{1}{3}\Lambda}{k^{2}}\right)^{-1}\right]\label{eq: betaGB}\end{equation}
and \[
\beta_{C}=\frac{1}{2}(2-\eta)\left[\frac{137}{60}+\frac{21}{40}\left(1-\frac{\frac{2}{3}\Lambda}{k^{2}}\right)^{-1}-\frac{9}{40}\left(1-\frac{\frac{4}{3}\Lambda}{k^{2}}\right)^{-1}-\frac{479}{360}\left(1+\frac{2\Lambda}{k^{2}}\right)^{-1}+\right.\]
\begin{equation}
\left.+\frac{269}{360}\left(1+\frac{\frac{4}{3}\Lambda}{k^{2}}\right)^{-1}+\frac{179}{45}\left(1+\frac{\Lambda}{k^{2}}\right)^{-1}+\frac{59}{90}\left(1+\frac{\frac{1}{3}\Lambda}{k^{2}}\right)^{-1}\right]\label{eq: betaC}\end{equation}
Now, everything depends on the expression for the anomalous dimension
$\eta$ of the TT graviton field. Using its expression motivated by
perturbative one-loop considerations, we get that
\begin{equation}
\eta=-\frac{1}{\omega_{C}}k\frac{d}{dk}\omega_{C}(k)
\end{equation}
where the coupling $\omega_{C}$ is related to the original Weyl coupling
by the inverse relation (i.e. $\omega_{C}=\alpha_{C}^{-1}$). This
extension to include a non-trivial anomalous dimension is here for
free since it does not change anything related to the structure of
the two beta functions that we have obtained.

To check the consistency of these results for the beta functions $\beta_{C}$
and $\beta_{{\rm GB}}$ one can obtain their asymptotic values in
the UV asymptotic regimes of energies. This means that we should take
the limit $k\to+\infty$ together with $\eta\to0$ since this anomalous
dimension is a quantum effect of higher order compared to the perturbative
one-loop running that the true running asymptotes to in the deep UV
regime. These limits result in the following expressions\begin{equation}
\beta_{{\rm GB}}=-\frac{87}{20}\end{equation}
\begin{equation}
\beta_{C}+\beta_{{\rm GB}}=\frac{137}{60}\end{equation}
 which are solved by \begin{equation}
\beta_{C}=\frac{199}{30}\quad{\rm and}\quad\beta_{{\rm GB}}=-\frac{87}{20}.\end{equation}
These two results perfectly agree with the one-loop perturbative results
as obtained earlier by Fradkin and Tseytlin.

Having obtained the results for the non-perturbatively improved beta
functions with the inclusion of the threshold effects of massive modes
and of anomalous dimension of the graviton $\eta,$ we can now discuss
the implications for the RG flows and its topology in QWG. The situation
in the UV regime is quite trivial and boring since in QWG both couplings
reach a trivial UV FP of RG. The coupling $\beta_{{\rm GB}}$ is induced
by quantum corrections to the action of QWG theory in (\ref{eq: weylgravaction}),
but as we have seen in (\ref{eq: GBvar}) it non-zero value does not
lead to any violation of conformal symmetry even in the local version.
This means that the perturbative couplings that should be defined
as $\left(\sqrt{\omega_{C}}\right)^{-1}=\sqrt{\alpha_{C}}$ and $\left(\sqrt{-\omega_{{\rm GB}}}\right)^{-1}=\sqrt{-\alpha_{{\rm GB}}}$
reach an asymptotically free FP in the UV, when the FP values of these
couplings both vanish. (The coupling of the Gauss-Bonnet term is negative
since its beta function $\beta_{{\rm GB}}<0$ is also negative and
this is why we took the definitions with the change of the sign under
the square roots.) We conclude that in the UV the quantum theory of
Weyl gravity meets a Gaussian FP with trivial scaling of couplings
and operators. This behavior is in agreement with that in the UV regime
the QWG theory should meet a conformally invariant FP where this model
is probably merged to a broader theory such as CSG in which the UV
FP with fully realized quantum conformality is present. In this UV-complete
theory there is no running and the value of the Weyl coupling can
be arbitrary (even non-perturbatively big), but still with any value
of $\alpha_{C}$ the conformal symmetry is explicitly present and
not violated by quantum effects.

Instead, the situation towards the IR regime of the energy spectrum
is more interesting. One looks in that regime for non-trivial new
FP of RG flows. They are defined by the simultaneous condition that
$\beta_{C}=0$ and $\beta_{{\rm GB}}=0$. One can see that for this
system of equations the presence of the anomalous dimension $\eta$
in the expressions for the beta functions (\ref{eq: betaGB}) and
(\ref{eq: betaC}) is completely spurious and this higher order quantum
effect can be safely neglected here. The results we have found are
really curious since for some finite value of the energy (in the IR
regime) we see with a very good precision of 2\% a non-trivial point
where the RG flow for the two beta functions $\beta_{C}$ and $\beta_{{\rm GB}}$
stops completely for a moment (of energies, or a while in RG time
$t=\log\frac{k}{k_{0}}$). This confluence of two FP's for respectively
$\alpha_{C}$ and $\alpha_{{\rm GB}}$ couplings is a very stunning
feature that is present only in QWG. Other higher derivative quadratic
gravities do not enjoy such a behavior in the IR regime. When one
uses a rescaled dimensionless energy RG scale variable, here conveniently
defined as \begin{equation}
\kappa=\frac{k}{\sqrt{|\Lambda|}},\end{equation}
then one can numerically determine the location of these FP's for
both of the couplings separately. First, in the case of $\Lambda>0$
we have a very good agreement and the FP for the $\alpha_{C}$ coupling
is found around $\kappa_{C}\approx1.17709$ and for the $\alpha_{{\rm GB}}$
coupling is found around $\kappa_{{\rm GB}}\approx1.19163$. For the
case of $\Lambda<0$ we still find a good agreement and the particular
values for energy locations of FP's are $\kappa_{C}\approx1.49722$
and $\kappa_{{\rm GB}}\approx1.52128$ for the couplings $\alpha_{C}$
and $\alpha_{{\rm GB}}$ respectively.

Now, we provide some interpretation to this special point (happening
almost at the same energy scales for both couplings) at some finite
energy scale. It cannot be a true IR FP of RG flow since this happens
at finite energy scale, so the flow may momentarily stop, turn back
but generally must continue, when we decrease the value of the RG
time $t$ more towards the deeper IR regime. We decided to call such
a special point of RG flow as turning point (TP) in distinction to
a true IR FP. One can plot the beta functions of the two couplings
analyzed here. One sees that the curves for RG beta functions cross
zero line almost at the same energy scale. Moreover, one expects that
by the inclusion of higher order phenomena (like by going to the two-loop
level) one will improve the accuracy of matching such that in a truly
non-perturbative theory the two curves cross zero exactly at the same
RG time. We can call this position as a $t_{{\rm TP}}$. The estimates
that we have made above suggest that its value is around $1.2$ and
definitely it is for values bigger than $1$. We emphasize once again
that this TP is not a true FP of RG that still has to be searched
for in the deeper IR regime. Instead, for this turning point there
exists a holographic interpretation that we describe in short below.

When one wants to describe the IR regime, one leaves the turning point
and continue the RG flow still towards the IR regime. Here we can employ
the perturbation calculus in value of the couplings that we met at
the TP. We remind that only the beta functions of couplings vanish
there, while this is not the case for the coupling themselves. We
denote a general coupling there by $\omega_{*}$ at the location of
TP. We therefore later perform perturbations in the small $\omega_{*}$
parameter treating the TP as a starting point for this perturbation
and assuming that $\omega_{*}\ll1$. The consequences of this are
the following. We found a non-trivial IR FP at $k=0$ so in the true
IR regime. This IR FP of RG is completely stable in a sense that the
dimensions of two operators $C^{2}$ and ${\rm GB}$ near it are finite
negative and bounded. Their actual values are equal to each other
and given by $\theta=-\frac{1}{3}$. This implies that they are relevant
operators near this IR FP. They describe the RG flow of the two originally
dimensionless couplings $\alpha_{C}$ and $\alpha_{{\rm GB}}$ of
the theory. One sees that thanks to the quantum effects the operator
and couplings acquire some anomalous but controllable dimensions and
in the IR regime they are not anymore dimensionless (and the corresponding
operators $C^{2}$ and ${\rm GB}$ are not of energy dimension $4$
in $d=4$). This means that the perturbation by addition of these
two operators to the CFT of IR FP of the theory is completely stable
and does not destabilize the physics which in the deep IR regime is
controlled and dictated by IR FP of the theory. Moreover, one finds
the following relation between the couplings $\omega_{*}$ defined
at the TP and the ensuing couplings $\omega_{**}$ at the true IR
FP, \begin{equation}
\omega_{**}=\omega_{*}+\frac{9}{2}\kappa\beta'(\kappa),\label{eq: relation}\end{equation}
where $\kappa$ is the schematic notation for the location of the
TP, and also the derivatives of the beta functions of the corresponding
couplings are evaluated at the TP location. This formula is valid
for both types of couplings $\omega_{C}$ and $\omega_{{\rm GB}}$.
Additionally, one finds that the couplings $\omega_{*}$ are not constrained
at all and the couplings $\omega_{**}$ are only forced to stay in
relation (\ref{eq: relation}), where the couplings $\omega_{*}$
are completely arbitrary. This means that also resulting true IR FP
values of the couplings $\omega_{**}$ can be both in the perturbative
as well as in some other non-perturbative regimes.

\section{Interpretation and further discussion of the results}

Firstly, we would like to discuss the possible interpretation of our
findings about the TP of RG flow that occurs at some finite location in
energy scales (around $\kappa=1.2$, so for $k=1.2\sqrt{\Lambda}$
for $\Lambda>0$). The whole fact that this is a turning point of
RG corresponds to the existence of a multi-branch holographic RG
flow. A possibly correct analysis of such an IR behavior
is provided by the AdS/CFT correspondence, which describes the geometrized
understanding of RG flows in various QFT's ( see ~\cite{Fukuma:2002sb} and references within). The geometrization comes
by promoting RG flows to some gravitational spacetime (characterized
by some metric tensor) in some higher dimensional spacetimes which
resembles the deformations of exact AdS spacetime. In our case, this
extended completely artificial spacetime is 5-dimensional and has
nothing to do with real physical gravitational spacetime considered
as exact or background solution in CWG. The geometrization of the
RG flow is only a useful theoretical tool to visualize various properties
of the otherwise complicated RG flows in quantum field theories \cite{holo}. As
the name suggests only in the case of FP of RG (when we know that
they are described by some CFT's) we are sure that the resulting gravitationally
dual 5-dimensional bulk is exactly identical to some patch of the
anti-de Sitter spacetime.  The departure
from the FP is encoded on the dual side of the theory by some radial
geometry that differs from the AdS. Therefore in the case of the RG
running of couplings we need to speak about the modified AdS spacetime
used for geometrization of RG flow. This bulk spacetime can be also
understood as a gravitational dual in 5-dimension to the field theory
which was considered in $d=4$ dimensions of real physical spacetime.
When the running takes place only as a function of one variable, like
in our case, where $k$ is its argument, then in the 5-dimensional bulk
we can use a metric of 5-dimensional spacetime which has a very special
metric structure. Namely this could be a conformally flat metric (in
a 5-dimensional sense) with only a non-trivial scale factor and in
this way this analogue of the FLRW spacetime used very frequently
in description of physical 4-dimensional cosmology. This is the framework
and the metric ansatz with which we will try to describe the geometrization
of the interesting RG flow phenomena that we have found towards the
IR regime in QWG. In particular, the situation with TP of RG is well
described by bounce cosmological solutions.

One recalls that this TP happens at some finite energy scales $\kappa$.
They in general, correspond to some finite radial locations in the
asymptotically AdS spacetime, where we geometrize the RG flow. Of
course, due to the fact that both in the UV regime as well as in the
IR we find a FP of RG in QWG, the gravitational dual spacetime must
be asymptotically (that is for zero and infinite values of the radial
locations) approaching some 5-dimensional AdS types of spacetimes.
Precisely, in the dual gravitational spacetime, the IR FP corresponds
to infinite radial location $\rho\to+\infty$ and UV FP is in turn
located at the origin of the AdS spacetime, that is for $\rho\to0$.
In general, there exists a relation between the energy scale in RG
flow $k$ and the radial coordinate $\rho$ of the dual AdS, which
reads $\rho\sim k^{-1}$, which is true up to a constant of proportionality.
The particular spacelike location of the TP of RG flow is dual to
4-dimensional surface (so co-dimension 1 in 5-dimensional bulk spacetime)
embedded in AdS-like 5-dimensional geometry, which is located at particular
constant value of the radial coordinate $\rho={\rm const}$ of the
AdS-like geometry. From this setup one easily understands that the
true IR FP of RG flow must corresponds to the conformal boundary of
AdS spacetime which is asymptotically located at infinite values of
the radial coordinate $\rho$. Therefore the TP cannot be a true IR
FP as we have also found earlier.

One can consider the gravitational cosmological spacetimes describing
the bounce situation. A particular useful example is the case of a
FLRW spacetime with a specific form of the scale factor $a(t)$. In
cosmology if the scale factor is not an injective function but a continuous
function of the cosmological time $t$, then there must exist a bouncing
point. This moment of the cosmological time corresponds to the physical
bounce like in the motion of a material point in the given gravitational
potential, when it bounces from the floor. This means that the scale
factor just after the bounce grows in time and just before it must
decrease in time. Such a gravitational spacetime is not very common
in cosmology since a bounce cannot be caused by the presence of normal
matter (satisfying the energy condition). The gravitational source
for the bounce must be an exotic matter. Typically if the bounce has
to take effect just for a very short period of cosmological time $t$,
then this non-physical source of matter must be localized and put
on the 4-dimensional brane located at some specific location in radial
coordinate $\rho={\rm const}$. In different vein, one can see this
gravitational spacetime as solution in Einsteinian gravity, where
the energy matter source is solely determined due to the presence
of some bulk scalar field with some non-trivial spacetime profile.
The fact that the bounce is sourced by exotic matter corresponds to
the fact that the scalar field must come with a negative kinetic term
(so then this is opposite to the standard scalar field and the scalar
field is of the phantomic character). In particular, one can take
the scalar profile to be only radially dependent, while in other transverse
dimensions it can be completely translationally invariant. Then with
such a bulk scalar field the bounce situation can be easily realized.
Actually by Friedmann equations of the ensuing cosmological model,
the profile of the scalar field is related to the beta function in
the RG flow. The TP of the RG flow is the situation when the beta
function from both sides of the running stops. This is a non-analytic
behavior in coupling when it is written as a function of the running
energy scale $k$. Such a behavior happens of course because the TP
is at some finite energy scale. The profile of the dual scalar field
in the 5-dimensional bulk spacetime must in such circumstances also
show some singularity. The detailed analysis shows that in the generic
case (when $\beta'(\kappa)$ is non-zero so the zero of the beta function
is a single, not multiple zero) the singularity of the scalar field
profile must be of the type of a square-root like singularity $\sqrt{x}$.
And obviously, such a behavior is non-analytic in the energy scale
$\kappa$.

The introduction of the scalar field is one of the way to interpret
geometrically the RG flow which exhibits the TP in the IR regime,
but for some finite energy scale. The non-analytic behavior in the
solution for the profile of the bulk scalar field can be analyzed
also from a different point of view. The TP is a critical point of
the RG flow and it corresponds to the joining (branching point) of
the two real solutions for the profile. This is seen clearly from
the type of the singularity which is here $\sqrt{x}$, so it describes
two real solutions which are exactly non-analytic at the point of
joining. This is interpreted as a very interesting situation for the
bulk scalar field, when the profile is not analytic function of the
radial coordinate $\rho$ of the AdS-like spacetime. The TP describes
then precisely a bifurcation point in the theory of scalar field on
this spacetime, where the two real solutions are joined. Each of the
two real-valued solutions describes a different RG flow, a different
branch of the RG flow. Hence we conclude that using the holographic
dictionary we are able to find an interpretation of the TP of RG flow
as a bifurcation point of a two-branch holographic RG flow in the
5-dimensional bulk spacetime.

Let us summarize what were our findings about the IR behavior of the
QWG. First, we found a TP (turning point) for RG flow at some finite
energy scales. These values of energies were moreover dependent on
the $\Lambda$ parameter which conveniently describes the curvature
scales on a general ES. However, we also found perturbatively a true
IR FP happening at $k=0$, so in the deep IR regime. This last true
IR FP fulfils all the requirements for the scale-invariant and terminal
point of the evolution of couplings described by RG flows. There formally
we have $k=0$ as the IR energy scale (or the limit $t\to0$ in the
RG time coordinate). It should be described by a CFT with gravitational
interactions in the infrared regime. Hence the evidence for such a
FP that we have collected, is quite remarkable and give rise to a
new FP with quantum gravitational interactions which also enjoy conformal
symmetry.

The TP that we found for some finite energy scales (of the order of
the curvature on ES) has origin that clearly can be explained. It
is there entirely due to the threshold phenomena related to the decoupling
of massive modes present in the theory around a general ES backgrounds.
Another non-perturbative phenomena is related to the inclusion of
an impact of the anomalous dimension $\eta$ of the quantum graviton
field. But as we have seen it does not change anything related to
the quest for IR FP's of the RG flows, even when its supposed truly
non-perturbative exact behavior is used in the computation. We find
that the TP happens for any initial values of the two couplings $\omega_{C}$
and $\omega_{{\rm GB}}$ that could be set at some UV energy scale
$t_{{\rm UV}}$; we did not see any constraint for them and they remain
completely arbitrary. But still the location of the TP of RG was determined
unambiguously and independently on these initial values for the RG
flow. As for what regards the IR FP (true one at $k=0$) we derived
only a relation (\ref{eq: relation}) which still leaves some arbitrariness
in the FP values of the couplings $\omega_{C**}$ and $\omega_{{\rm GB}**}$.
We see that there is no any constraint from which we could possibly
determine the special values of the coupling constants $\omega_{C*}$
and $\omega_{{\rm GB}*}$at the TP and hence also the values of $\omega_{C**}$
and $\omega_{{\rm GB}**}$ at IR FP. Therefore we conclude that in
the IR FP's of QWG we are on a 2-dimensional surface of FP's (which
is of course a simple generalization of a line of FP). This is indeed
considered as a very special situation for the IR behavior of the
quantum field theory. But here all the speciality comes because of
the special character of quantum gravitational theory with conformal
interactions.

One can also analyze the applications of the IR behavior for the actual
physical cosmology of 4-dimensional Universe. For this we need the
Universe to be classically described by conformal gravitational theory.
The quantum effects captured by RG flow phenomena may have some interesting
implications and features for inflationary cosmology. First, one can
identify the UV FP of QWG as the moment when the characteristic energies
of all particles are very high, so this is a moment of the Big Bang
for our Universe. However, in this paper we put more emphasis on the
IR behavior due to quantum RG effects. One knows that near the IR
FP the symmetries of the theory may change and the theory undergoes
a phase transition. The same could happen with the Universe. Therefore
the IR FP may be an onset for the inflation where the dynamics of
the quantum gravitational field changes completely and start a new
phase with exponential expansion and with exponential growth of the
cosmological scale factor $a(t)$. One could also ask for the interpretation
of the scale $\Lambda$ which was crucial for a determination of the
location of TP of the RG flow. As we defined this was a characteristic
scale of the curvature on a general ES, in particular related to the
radius of curvature in the case of MSS. In the $4$-dimensional cosmological
framework the MSS are realized as physical de Sitter spacetimes characterized
by some radius describing actually the speed of the exponential expansion.
Hence, when $\Lambda>0$ one sees that the de Sitter background that
we use in our computation may be interpreted as inverse radius (up
to some constant of proportionality) of the inflationary phase of
the Universe. In this way our theoretical considerations related to
the RG effects may have some direct interpretations and implications
for physical cosmology of our Universe.

One of the two pending problems of QWG is the issue of conformal anomaly
(CA)~\cite{Capper:1974ic,Duff}. Here we will comment on it in more details. The other problem
is related to the presence of Boulware-Deser ghosts (or also known
as Weyl ghost in Weyl gravity) which quite likely endanger unitarity
of the theory when it is quantized in a standard way. Therefore the
quantization procedure of classical Weyl gravity probably requires
some modifications. We will not dwell on this interesting issue with
apparent violation of unitarity in QWG anymore. Instead we discuss
some issues with CA.

The trace anomaly has to do with the trace of the energy momentum
tensor of the physical system in question. When gravity is coupled
to matter then for this issue we have to consider the whole system
(gravity and matter). For such a system we define the classical trace
of the total energy momentum tensor (EMT)\begin{equation}
T=g_{\mu\nu}T_{{\rm tot}}^{\mu\nu}\quad{\rm or}\quad T=\left\langle \hat{g}_{\mu\nu}\hat{T}_{{\rm tot}}^{\mu\nu}\right\rangle \end{equation}
for the quantum version of the theory when we have to deal with quantum
operators and their expectation values. The energy momentum tensor
of the total system we obtain as a variational derivative of the total
action of the system with respect to metric fluctuations, that is
we have\begin{equation}
T_{{\rm tot}}^{\mu\nu}=\frac{2}{\sqrt{|g|}}\frac{\delta S_{{\rm tot}}}{\delta h_{\mu\nu}}.\end{equation}
One finds according to Weinberg definition and understanding of the
total EMT \cite{Weinberg} that on-shell (so using gravitational equation of motion),
this tensor for the whole system vanishes. One could interpret this
as the fact that the EMT of matter completely balances the one coming
from gravitational sector. On the classical level one finds that off-shell
the EMT is non-zero of the total system. This also implies that off-shell
the trace $T$ in general theories is not vanishing.

Let us consider the off-shell trace $T=T_{\mu}^{\mu}$ of the total
EMT of the system read from the total action (could be classical tree-level
action or quantum effective action). If in the gravitational sector
the action is given by\begin{equation}
S=\!\int\! d^{4}x\sqrt{|g|}R^{2},\end{equation}
which is a diffeomorphic invariant and if this is the total system
(no matter present), then\begin{equation}
\frac{2}{\sqrt{|g|}}\frac{\delta S}{\delta g_{\mu\nu}}=T_{{\rm tot}}^{\mu\nu}\neq0\end{equation}
and this is off-shell non-zero, so off-shell the trace $T\neq0$ in
general. Only for the Weyl action (\ref{eq: weylgravaction}) we have
that off-shell\begin{equation}
T=g_{\mu\nu}\frac{2}{\sqrt{|g|}}\frac{\delta S}{\delta g_{\mu\nu}}=g_{\mu\nu}B^{\mu\nu}=0\end{equation}
as the result of special geometric properties of the Bach tensor.
We look for a similar speciality on the quantum level of gravitational
theories with conformal symmetries present not only on the classical
level.

The special situation happens in classically conformally invariant
theories (both with gravity or just of pure matter). Then one finds
that the trace of EMT even \emph{off-shell} vanishes identically and
this is a clear sign of conformal invariance of the total classical
theory. Of course, the total trace vanishes on-shell but this is due
to the fact that on-shell $T_{{\rm tot}}^{\mu\nu}=0$, so the condition
$T=0$ on-shell does not say nothing about the conformal invariance.
By closer inspection one notices that in some models of matter which
are specially non-minimally coupled to gravity, the following
complication arises. One sees that if the action of the total theory
is conformally invariant, then the trace of the EMT of the total system
vanishes but only with using of matter sector EOM. But still the gravitational
sector EOM are not used to find that $T=0$, so this is mixed off-shell
and on-shell situation, but it is important that for the main degrees
of freedom for EMT, that is for metric fluctuations the situation
is off-shell. To derive any conclusion about conformal invariance
(in the GR sense) we must not use gravitational EOM, but it is possible
to use the ones from the matter sector, or it is even necessary. The specific
models in question for which such situation happens are for example
conformally coupled scalar fields in $d=4$ with the action\begin{equation}
S_{\phi}=\!\int\! d^{4}x\sqrt{|g|}\phi\left(\square+\frac{R}{6}\right)\phi\end{equation}
or also higher derivative conformally coupled Dirac fermions and gauge
vectors in higher number of dimensions. In this way we can easily
derive about the conformal invariance in the theory on the classical
level. And this is indeed verified by explicit computations for all
known classical conformal models.

On the quantum level if we want to preserve conformality, then we
should again require that $T=0$ (or even in the sense of expectation
values $\left\langle T\right\rangle =0$). This is also the condition
for the absence of CA on the quantum level. Known models with fully
preserved quantum conformality satisfy this condition. Now, the
pertinent question is whether there is or there is not CA in pure
QWG theory analyzed by the means of the standard quantization procedure.
We should accept that this must be viewed as a truly off-shell problem
from the point of view of gravitational EOM. It is interesting also
to study in details the situation when matter models are coupled in
a conformal way to CWG gravity. Historically it was first that the
CA appeared as generated due to matter loops in matter models that
were put on curved external gravitational backgrounds. It was the
inconsistency between the classical conformal symmetry of the matter
models (like for example electrodynamics with massless fermions) and
the breaking of it on the quantum level, already for one-loop level
for example. Due to a presence of non-vanishing beta functions of
classically dimensionless couplings, or in general due to RG flow,
the anomaly shows up. Only in some special models like in ${\cal N}=4$
supersymmetric Yang-Mills theories, the CA was cancelled and there
were no beta functions, and in the result the theory was UV-finite~\cite{Brink:1982wv}.
There was a hope that in matter models with CA the coupling to gravity
will balance contributions to beta functions and the problem of CA
would be solved in this way.

However, explicit computations of CA, performed even to the simplest
one-loop level, showed us that this is generally impossible. The conformal
anomaly shows up in particular in the computation of triangle diagrams
that we have at the one-loop level in a general matter theory. One
could also notice a very interesting observation that at the one-loop
level in $d=4$ spacetime dimensions the contributions to anomaly from
gravitational and matter sector are completely independent one of
each other and the total anomaly is the sum of all of them without
any interference terms. This is originated from the expressions for
the beta functions of gravitational terms $R^{2}$, $C^{2}$ and ${\rm GB}$,
which are linear in the contributions coming respectively from the
matter sectors and gravitational sector. Therefore if there is a balance
in the total anomaly, then this is achieved without any participation
of mixing terms. But this also implies that the CA in pure QWG cannot
vanish since matter is needed for the balance and the linear superposition
is at work here. Therefore it is very crucial to consider coupled
matter plus gravity models, when the balance of these two contributions
is theoretically possible. The example of coupled theory is ${\cal N}=4$
CSG with two copies of the ${\cal N}=4$ SYM matter species, where
the CA completely vanishes as this theory is called as anomaly-free
by its authors (Fradkin and Tseytlin).

The fact that on-shell the trace $T$ of the total energy-momentum
tensor of the system vanishes \cite{Mannheim:2011ds} has implications only for anomaly of
global conformal symmetry. By adding a spin-2 field (understood as
matter not as a dynamical spin-2 geometrical current), coupled to
external geometry represented here by the tensor of rank-2 namely
the metric of the spacetime, one can cancel the global anomaly ('t
Hooft anomaly) of the conformal invariance. This is evident from the
fact that on-shell the energy-momentum tensor of the total system
(gravity+matter) vanishes $T_{\mu\nu}=0$ \cite{Weinberg}. For the cancellation of
the global anomaly this condition is exactly necessary and sufficient
and should hold on-shell. Similarly, a gauge current conservation
in some globally gauge anomalous theories (because of the presence
of chiral fermions) is due to on-shell condition and after all (after
adding some new matter fields) the total current is conserved, so
there is not a global anomaly anymore. For the 't Hooft global anomaly
(and current conservation or the issue of the trace of the EMT) the
off-shell situation does not matter at all.

We agree that on-shell we have $T_{\mu\nu}=0$ for the total system
(gravity+matter). But this does not cancel the anomaly of the local
gauged conformal symmetry which is supposed to be a local symmetry in conformal
gravity. Analogously, $\partial_{\mu}j^{\mu}=0$ does not guarantee
that local gauge anomaly is cancelled in models with chiral fermions
and gauge symmetries. It would only guarantee that global gauge anomaly
is not there in the model. 
But for local gauge anomalies if they are
there the theory is sick and should not be considered on the quantum
level. Moreover, the argument from \cite{Mannheim:2011ds} about vanishing of the total
EMT on-shell seems too much robust and holds for any gravitational
theory (with diffeomorphism symmetry) and with any conformal matter
as a gravitational source (so this matter must have $T_{m}{}_{\mu}^{\mu}=0$
on matter EOM to be conformally coupled). The original problem with
CA was originated in matter sectors, and this was global conformal
anomaly. Some people claim that the anomaly from the matter side is cancelled
by coupling to conformal gravity. But pure conformal gravity should
solve this problem by its own. So how is it that the contribution
of gravity is enough for gravity alone, but adapts to the matter in
such a way that this reaches a balance and the total contribution to
CA is zero off-shell? It is impossible, if one knows that the beta
functions at the one-loop level are independent and additive (for
two sectors: matter and gravity), but here the contributions do not
adapt to whether there is matter added or not to always balance, so
the total anomaly does not vanish in full generality.

Moreover, that argument seems to be too naive, because for any diffeomorphism-invariant
gravitational theory on-shell for the total system $T_{\mu\nu}=0$.
For this to happen, this does not have to be a conformal gravity.
And then automatically $T=0$ on-shell, so there is not a conformal
anomaly in any gravitational theory, no matter if conformal or not,
or whether it is coupled to any matter sector or not. This seems to
be too robust argumentation. But we think that the conformal symmetry
should put some constraints, like it puts in Fradkin Tseytlin CSG.
And then you have to add a very special matter sector to QWG to make
it anomaly-free.

As we emphasized in the introduction if we have that in the quantum
model $T\neq0$, then we generally run into problems in quantum theory.
One sees that the classical theory (before quantization) was with
local conformal symmetry, while quantum theory is without it. There
is no enough symmetry on the quantum level to bring the positive features
of conformality which is very constraining as we have seen before.
This symmetry is for example needed to constrain the form of all possible
UV-divergences (or the absence of them in completely UV-finite theories \cite{LM})
and also is essential to restrict very much the form of the quantum
effective action in such a theory on every loop level and also non-perturbatively.
If one understands that this symmetry is not present on the quantum
level, then the problems are with Green functions which now do not
respect the full symmetries of the classical theory. This means that
quantum effects break the gauged conformal symmetry of the model -
the symmetry is not there on the full quantum level and conformal
Ward identities are not satisfied. The theory on the quantum level
is not very much controlled anymore. One basically ends up with quantum
version of the theory quadratic in gravitational curvatures (a general
higher derivative Stelle theory) and the traces of conformal symmetry
which was present only on the classical tree-level are gone. The quantum
theory is not special anymore. Of course, such situation is very bad
and should be avoided by all means. So, in the sense CA in the UV
regime should be made vanishing.

When one analyzes the pure $C^{2}$ gravity, so the Weyl conformal
gravity in $d=4$, one finds that there is a non-trivial RG flow that
we exploited in this paper. The beta function of the Weyl coupling
$\beta_{C}$ does not vanish even at the level of first quantum loop.
For higher loop this situation is even more dangerous. One could be
misled by this fact since this is the UV-divergence proportional to
the conformally invariant counterterm $\sqrt{|g|}C^{2}$ in the UV-divergent
part of the effective action. Somehow, this situation knows about
the remnants of conformal symmetry since counterterm is conformally
invariant, however just the mere fact of its presence signifies that
there is a non-vanishing CA and that there are severe problems with
the preservation of the conformal symmetry on the quantum level. Moreover,
one also remembers that on the one-loop level there is a non-vanishing
beta function of the Gauss-Bonnet term. However, one can waive this
counterterm since its variation (to get for example quantum EOM or
any contribution for Green functions) is a total derivative and it
is in particular conformally invariant. The counterterm which would explicitly
break the conformal symmetry in a full local version, namely the $R^{2}$
counterterm is not present there on the level of one-loop. However,
it is heralded due to the presence of CA already on the one-loop level
and one expects the need for presence of the $R^{2}$ counterterm
starting from the two-loop level on. One can also dismiss these arguments
with beta functions since they are not directly related to any observable
effects, and the RG running of couplings is only a theoretical effect
predicted by theory. In general, it is very difficult to find an
observable consequence of the violation of the conformal symmetry
on the quantum level explicitly such that everyone would agree with
it unanimously. Similarly the trace of the quantum EMT of the total
system is difficult to measure directly and one does not have any
direct physical phenomena derived from the breaking of conformal symmetry.

In this quite disappointing situation, one has to look at other objects
that still have good physical meaning. One of such an object is the
quantum effective action, in particular not only its UV-divergences
but also finite terms. Therefore it is very useful to analyze the
situation with some selected terms of the quantum effective action.
We choose some of them from the infinite number of them that they
are expected in the effective action (even restricted to the one-loop
level) due to the infinite series expansion in number of derivatives
and another series expansion in powers of fields present in some terms
of it. The analysis presented below is more transparent regarding
the issues of conformal symmetry than the analysis with UV-divergences
at the one-loop level since as written above we found that all counterterms
generated at the one-loop level are conformally invariant. We will
see the explicit violation of conformal symmetry from some terms in
the effective action.

Due to the RG invariance of the total effective action $\Gamma$ and
the existence of the non-vanishing beta function $\beta_{C}$, one
derives that in the finite terms of the effective action at the one-loop
level one must have a term\begin{equation}
\Gamma_{{\rm fin}}\supset\beta_{C}C_{\mu\nu\rho\sigma}\log\left(\frac{\square}{\mu^{2}}\right)C^{\mu\nu\rho\sigma},\end{equation}
where $\mu$ is an arbitrary renormalization scale needed to absorb
the UV-divergences of the theory. One could say that just the fact
that we have to introduce this dimensionful parameter $\mu$ to make
sense of the theory (i.e. to renormalize it) explicitly breaks the
conformal invariance of QWG. But one could wait with this argumentation
by saying that $\mu$ is a spurious non-physical parameter that all
observable measurable effects should be independent of, so about its
existence we derive only theoretically, while experimentally it is completely
invisible. None can polemize with this fact. However, our argumentation
goes one step further. And we consider the full quantum action (classical
tree-level action of the QWG theory and the term with first quantum
corrections at the one-loop level). Such an object in $d=4$ reads
explicitly\begin{equation}
\Gamma_{{\rm tot}}=\Gamma_{C^{2}}+\Gamma_{{\rm eff}}\supset\!\int\! d^{4}x\sqrt{|g|}\left[\alpha_{C}C^{2}+\beta_{C}C_{\mu\nu\rho\sigma}\log\left(\frac{\square}{\mu^{2}}\right)C^{\mu\nu\rho\sigma}\right].\label{eq: totaction}\end{equation}
The formula above contains some selected terms in the full effective
action to the one-loop level with all quantum corrections taken into
account. But one should view this functional as giving the action
description of the theory to the one-loop level with full quantum
corrections included. So one should treat it as the classical action
and one could in principle derive all correlation functions and scattering
amplitudes from it to the level of one-loop accuracy. But if this
functional should be seen as classical one could also answer the question
about the presence of symmetries on the one-loop quantum level, in
particular about the presence or absence of the conformal symmetry.
This question is equivalent to asking whether the \emph{classical}
theory governed by the action (\ref{eq: totaction}) is conformally
invariant. One can even simplify his life by neglecting the presence
of the scale $\mu$ there and we can be silent about its conformal
transformation law. This will be immaterial for what follows next.
When one considers the action term\begin{equation}
\sqrt{|g|}\beta_{C}C_{\mu\nu\rho\sigma}\log\square C^{\mu\nu\rho\sigma},\end{equation}
then one sees that the only case in which this is conformally invariant
in $d=4$ is when $\beta_{C}=0$ (everything is spoiled here by the
presence of the box operator which transforms non-trivially and non-homogenously
under conformal transformations, and on top of that by the logarithm
function of such an operator which acts on a tensor representation
of Weyl tensor naturally appearing with its four indices). This term
in the total effective action is not conformally invariant in $d=4$
unless its front coefficient vanishes. The condition $\beta_{C}=0$
is then more direct condition for the vanishing of the CA on the quantum
level. The terms in the effective action are more related to observable
quantities than counterterms or beta functions, which serve here theoretically.
One concludes that in pure QWG there is CA, so there is a problem
with preservation of the full conformal symmetry on the quantum level.

Since we have confirmed that on the quantum level the conformal symmetry
is not fully realized in QWG, then one can ask for reasons of that
behavior. One explanation is due to the fact that on the quantum level
quantum polarization effects touch conformal symmetry in a dramatic
way and the results are very devastating since the conformal symmetry
is hardly broken there, when analyzed in the UV regime. Pure QWG theory
is anomalous which puts a big question about the consistency of this
theory since this conformal symmetry was gauged in such a model. This
is in simple words that we expected the symmetry to be on the quantum
level and we quantized a theory putting such a lot of care to treat
conformal symmetry delicately, but the result is opposite. The conformal
symmetry was there only on the classical, but it is not there anymore
on the quantum level. Of course, like we explained above this breaking
of conformal symmetry is clearly related to the RG running of couplings
and to non-trivial RG flow in general. However, we think that this
is a problem with the UV-completion of the theory, similarly like
the renormalizability was the problem with quantum Einsteinian gravity
before it was properly UV completed by terms with higher derivatives
in $d=4$ dimensions. If the UV-completion or UV-embedding of the
QWG theory is done by means of CSG, then the problem in the UV limit
is solved. One has to worry only about the problem of the CA in the
UV limit so for very high energies (when the notion of energy still
make sense). If UV-completion is achieved and there is not CA in the
fundamental embedding theory, then the problem of overall consistency
of the QWG model at lower energies is not an issue. One can also for
still lower energies study the RG flow and analyze the situation with
soft breaking of conformal symmetry in the domain of applicability
of QWG theory (so in the regime of energies when there is a running
of gravitational couplings). This is the direction that we have pursued
in this contribution.

Now, one can think about some of the remedies for the situation of
CA of QWG on the quantum level. Restricting to the first quantum loop
level, one can invent some ingenious solutions with adding proper
number and proper representations of matter species, in such a way
that in the linear superposition the total beta functions for the
dimensionless gravitational couplings like for $\alpha_{C}$, $\alpha_{R}$
and $\alpha_{{\rm GB}}$ are balanced to zero thanks to matter contribution
which act against the pure QWG gravity. This means that the special
matter content must be added to make the theory anomaly-free. Then
the dangerous non-conformally invariant terms like $C\log\square C$
are not generated in the quantum effective action. This could be viewed
as a provisory solution on the level of one-loop action, but these
new matter species can be promoted to fully gauge-invariant copies
of ${\cal N}=4$ SYM theories by subjecting to Noether procedure.
In this way quantum conformality will be present on the quantum level
of the full coupled theory. This will give back the possibility to
constrain the scattering amplitudes and expressions for Green functions
and various other correlation functions thanks to the presence of
quantum conformality. Finally, if on the full level of quantum effective
action the conformal symmetry is present in a preserved way, only
then one can successfully resolve the issues of GR-like singularities
of classical gravitational theories using the quantum effects from
the quantized gravitational theory and using the fully realized quantum
conformal symmetry. If this would be done only on the level of classical
action of WG, then one runs into the problem that quantum correction may
completely invalidate this conclusion about the resolution of singularities
and they may change drastically the quantum solutions of the theory
bringing back singularities. Instead, when one has a conformal symmetry
on the quantum level one is sure that resolution is final, stable
and that various quantum corrections cannot destroy it.

\section*{Acknowledgements}

We thank Petr Jizba, Roberto Percacci,  Ilya Shapiro and Philip Mannheim
 for valuable discussions on the topic of conformal gravity both in the classical
 and quantum domain. S.G. would also like to thank Loriano Bonora for relevant discussions on the issue of conformal anomalies. The research of S.G.\ has been supported by the Israel Science Foundation (ISF), grant No.\ 244/17.\\

\noindent {\bf References}\\

\end{document}